# Spinon mediation of witness spin dynamics in herbertsmithite

Hiroto Takahashi[1*], Jack Murphy[2*], Mitikorn Wood-Thanan[3,4*],
Pascal Puphal[5], Miguel-Ángel Sánchez-Martínez[4], Fabian Jerzembeck[1,6], Chun-Chih Hsu[1],
Jonathan Ward[2], Masahiko Isobe[5], Yosuke Matsumoto[5], Hidenori Takagi[5],
Stephen J. Blundell[1], Michael R. Norman[7], Felix Flicker[4**] and J. C. Séamus Davis[1,2,6,8**]

[1] *Clarendon Laboratory, University of Oxford; Oxford, OX1 3PU, UK*
[2] *Department of Physics, University College Cork; Cork, T12 R5C, IE*
[3] *School of Physics and Astronomy, Cardiff University; Cardiff, CF24 3AA, UK*
[4] *School of Physics, University of Bristol; Bristol, BS8 1TL, UK*
[5] *Max Planck Institute for Solid State Research; Stuttgart, D-70569, DE*
[6] *Max Planck Institute for Chemical Physics of Solids; Dresden, D-01187, DE*
[7] *Materials Science Division, Argonne National Laboratory; Lemont, IL 60439, USA*
[8] *Department of Physics, Cornell University; Ithaca, NY 14853, USA*
* These authors contributed equally to this work.
** Corresponding authors. Email: felix.flicker@bristol.ac.uk & jcseamusdavis@gmail.com

**The kagome lattice of spin-1/2 copper atoms in herbertsmithite is conjectured to sustain a quantum spin liquid state with spinon quasiparticles. Ideally, the kagome crystal planes are each separated by a plane of spinless zinc atoms. However, in real crystals some spin-1/2 copper atoms substitute randomly onto these inter-kagome zinc sites. Here we reconceptualize such 'impurity' atoms as quantum witness spins whose dynamics is designed to probe the spin liquid state. We then introduce spin noise spectroscopy to measure the frequency and temperature dependence of witness spin dynamics, demonstrating that their phenomenology is consistent with extensive interactions between witness spins mediated by propagation of spinons through a quantum spin liquid. Ultimately, a sharp transition occurs at around 260 mK, below which the properties of both spin noise and magnetic susceptibility suggest that the witness spins form a spin glass phase. Among theoretical models considered, we demonstrate that our observations are only consistent with spinon-mediated interactions between witness spins by either a $Z_2$ or U(1) quantum spin liquid, with the former model more closely matching the data. Our work demonstrates that quantum mechanical witness spins may now conceivably be used as a widely applicable probe of quantum spin liquid physics.**

In quantum materials research, 'impurity' atoms are usually viewed as pernicious. Yet this can sometimes be misguided. For example, zinc (Zn) 'impurity' atoms substituted at the copper (Cu) site in cuprate high-temperature superconductors provide a prolific source of fundamental understanding[1]. Here we deploy 'impurity' atoms as a resource for the direct atomic-scale study of a putative quantum spin liquid (QSL). In the theory of such materials, strongly interacting spins do not undergo spontaneous magnetic ordering, but enter a ground state with massive long-range quantum entanglement whose fractionalized charge-neutral quasiparticles are dubbed spinons[2-5]. Herbertsmithite, a kagome-lattice magnetic insulator with chemical structure $ZnCu_3(OH)_6Cl_2$, is a leading candidate to sustain such a QSL state[2-4, 6-8]. Moreover, theoretical analysis of the kagome Heisenberg spin-1/2 Hamiltonian[9,10] finds that such a QSL exhibits either a spectrum of delocalized spinons above a finite energy gap $\Delta$ or of a gapless nature, governed by $Z_2$ or U(1) gauge symmetry, respectively. Here, because in herbertsmithite, some spin-1/2 Cu atoms substitute randomly onto the inter-kagome Zn sites[7,8,11], our objective is to reconceptualize these 'impurity' atoms as 'witness spins' to provide an exceptional probe of the conjectured QSL state.

The herbertsmithite crystal structure (lattice parameters $a = b = 6.84$ Å ; $c = 14.09$ Å) reveals the kagome plane of $Cu^{2+}$ $s = 1/2$ atoms that are hypothesized to sustain a QSL (Fig. 1a). The contribution of these spins alone to the kagome-specific magnetic susceptibility, $\chi_{\mathrm{K}}(\omega, T)$, exhibits a strong downturn at $T \lesssim 30$ K, hypothetically due to spin-singlet formation[12,13]. At higher temperatures, the DC susceptibility $\chi(T)$ indicates an antiferromagnetic (AF) interaction scale $J_{\mathrm{K}} \approx 17$ meV between kagome spins, with a congruent Curie-Weiss temperature ($\theta_{\mathrm{CW}} \approx -300$ K) yet no magnetic ordering has been reported down to $T \lesssim 50$ mK[14,15]. As the leading QSL candidate, this material has been an intense focus of study, especially on whether the putative spinon spectrum is gapped or gapless. Some nuclear magnetic resonance studies report a spin-liquid ground state with a finite energy gap[16]; the low energy inelastic neutron scattering spectra are reported consistent with a small energy gap in the kagome planes[17]; and pressure tuning of the material reveals enhancement of the magnetic ordering temperature consistent with a gapped QSL state[18]. On the other hand, there also are nuclear magnetic resonance studies reporting that the kagome plane susceptibility at lowest temperatures reaches a nonzero value indicating the absence of an energy gap[19]; both specific heat and susceptibility studies report evidence that herbertsmithite is a QSL without a spin gap[20]; Raman scattering experiments provide further indications of a gapless QSL state[21]; as do optical conductivity experiments[22]; and specific heat measurements[23]. More recent nuclear magnetic resonance studies affirm strongly that the spin excitations in the kagome plane of herbertsmithite are gapless.[24]

There is, however, a second major reservoir of spins outside the kagome planes of herbertsmithite because substitution of $Cu^{2+}$ $s = 1/2$ ions occurs on approximately 33% of the $Zn^{2+}$ $s = 0$ sites[25] (Methods A). From this interplanar perspective (Fig. 1b), each kagome crystal plane separates two planes of $Zn^{2+}$ atoms (red) into which these $Cu^{2+}$ atoms (blue) substitute. Due to microscopically unexplained AF interactions between these $Cu^{2+}/Zn^{2+}$ substituted spins[14], the residual susceptibility as $T \to 0$ evolves approximately as $\chi^{-1}(T) \propto (T - \theta_{\mathrm{CW}})$ with an extremely low Curie-Weiss temperature $\theta_{\mathrm{CW}} \approx -1.1$ K. Equivalently, in $T \to 0$ neutron scattering studies, the higher energy structure-factor $\Sigma(\boldsymbol{q}, \omega)$ exhibits a broad continuum[17,26,27] possibly associated with spinons but, for $\hbar\omega \lesssim 1$ meV, $\Sigma(\boldsymbol{q}, \omega)$ is quite distinct and has been interpreted as due to AF interactions between $Cu^{2+}/Zn^{2+}$ substituted spins, driven by interactions thought to be transmitted through the kagome plane[17,28]. A major challenge to determining whether the kagome plane spin excitation spectrum is gapped[16] or gapless[24] has been to accurately quantify and understand physically the spin dynamics phenomenology of these $Cu^{2+}/Zn^{2+}$ substituted spins.

Although such substitutional 'impurity' spins are widely viewed as an impediment to analysis of the $T \to 0$ phenomenology of herbertsmithite[7,8] (Supplementary Discussion), here we exploit them as an innovative resource. Viewed as quantum mechanical 'witness spins', they may conceivably be used for exploration of the QSL state itself[24,29], through its influence on their spin dynamics. To explore this concept we introduce spin noise spectroscopy[30-34] to QSL studies, using a custom-built spin noise spectrometer (Fig. 1c) measuring the time evolution of flux $\Phi(t)$ generated by the $ZnCu_3(OH)_6Cl_2$ crystals. The apparatus uses a pair of opposite-chirality superconducting solenoids (Fig. 1c) in a continuous superconductive circuit including the input inductance to a DC-SQUID. The output voltage of the SQUID is $V_{\mathrm{D}}(t) \equiv \frac{\Phi_{\mathrm{D}}(t)}{\eta}$, where $\Phi_{\mathrm{D}}(t)$ is the flux delivered to the SQUID and $\eta = 0.10\frac{\Phi_0}{\mathrm{V}}$; the measured flux $\Phi_{\mathrm{D}}(t)$ relates to the flux threading the sample $\Phi(t)$ as $\Phi_{\mathrm{D}}(t) \equiv \beta\Phi(t)$ where for this apparatus $\beta = 0.0084$. Our spin noise spectrometer achieves magnetic field sensitivity $\mu_0\delta M \leq 10^{-14}\frac{\mathrm{T}}{\sqrt{\mathrm{Hz}}}$, on a cryogen-free ultra-low-vibration dilution refrigerator in the range 10 mK $\leq T \leq$ 5000 mK. The time-sequence of the magnetic flux $\Phi(t)$ generated by the sample magnetization $M(t) = \vartheta\Phi(t)/\mu_0$ within the pickup coil, where $\vartheta = 1.1 \times 10^{-10}$ T/$\Phi_0$, can be measured with microsecond precision. In a given experiment, typical measurables include $M(t,T)$, $S_M(\omega,T)$, $\sigma_M^2(T)$ and $\chi(T)$ [30-34]. Our $ZnCu_3(OH)_6Cl_2$ sample (Extended Data Fig. 1) preparation and evaluation procedure is described in Methods A.

Immediately upon commencing our herbertsmithite experiments, we discovered that, for $T < 400$ mK, all $ZnCu_3(OH)_6Cl_2$ samples begin to spontaneously generate robust spin noise. As shown in Fig. 2a, the spin noise at $T = 260$ mK fluctuates with maximum field-amplitudes $B(t) \equiv \mu_0 M(t)$ in the range of $10^{-12}$ T, orders of magnitude above the background. Figure 2b shows exemplary time-sequences of the measured magnetic flux $\Phi^2(t,T)$ for selected temperatures, demonstrating strong magnetization fluctuations that have Gaussian distributions (Extended Data Fig. 2a). From such data, we determine the power spectral densities of flux noise $S_\Phi(\omega,T)$, shown in Fig. 2b inset and Fig. 2c for 0.6 rad • Hz $\leq \omega \leq$ 600 rad • Hz, with equivalent magnetic field noise on the right axis. Even at this elementary stage, herbertsmithite haecceity appears striking, because fluctuations in the spin-1/2 magnetization of a mm-scale sample are spontaneously generating magnetic fields near $10^{-12}$ T. Finally, Fig. 2d presents a color-coded contour plot of measured $S_\Phi(\omega,T)$ revealing a clear transition in witness spin dynamics at $T^* \approx 260$ mK as indicated by the horizontal arrow. Since kagome spins have virtually no direct contribution to magnetic phenomena[12,13] at sub-Kelvin temperatures, and since the DC susceptibility $\chi(T)$ of our samples is quantitively consistent with the magnetization expected when approximately 33% of Zn sites are occupied by Cu, the observed spin dynamics can only be explained by witness spin contributions.

Fitting $S_\Phi(\omega,T)$ as in Fig. 2d, the spin noise spectra are found to be scale-invariant. The measured witness spin noise power-index $\alpha$ from fitting $S_\Phi(\omega,T) = A(T)\omega^{-\alpha(T)}$ (Fig. 3a) undergoes a clear transition to $\alpha \approx 1$ at $T^*$, indicated by the vertical line. Simultaneously, the measured witness spin noise variance $\sigma_\Phi^2(T)$, as determined by integrating $S_\Phi(\omega,T)$ in the range 0.6 rad•Hz $\leq \omega \leq$ 600 rad•Hz (Fig. 3b) also plainly indicates a transition in total noise power (vertical line). Thus the herbertsmithite witness spin dynamics undergo a sharp transition at $T^*$.

The witness spin magnetic susceptibility $\chi(T)$, measured in ultra-low magnetic fields $B \lesssim 4$ μT, reveals the well-known[14,35] Curie-Weiss behavior of $\chi(T)$ at temperatures $T \sim 1$ K, yielding quantitatively that the witness spin density is 32.5% $\pm$ 0.5% per plane of $Zn^{2+}$ sites in our samples (Methods A). However, we find that this diverging $\chi(T)$ is first diverted from precise $\chi^{-1} \propto (T - \theta_{\mathrm{CW}})$ and then interrupted by a sharp transition to a rapidly diminishing $\chi(T)$ below $T^*$, as shown in Fig. 3c. While no such transition has been observed previously using higher magnitude fields[14], we observe it in all samples (Extended Data Fig. 3). The combination of this cusp and noise transition suggests witness spin glass formation. Accordingly, we extract the Edwards-Anderson (EA) spin glass order parameter from these $\chi(T)$ data (Methods B). While zero above $T^*$, it increases rapidly below (Fig. 3d). To confirm witness spin glass formation, we monitor the evolution of the flux $\Phi(t)$ in a 2 μT applied field under sudden thermal quenches from a thermalized condition at $T_1 = 400$ mK to a lower temperature $T_2$ (Fig. 3d inset). Here, $\Phi(t)$ and thus the sample magnetization $M(t)$ enter a $-\ln t$ relaxation regime, requiring days to equilibrate, but this effect begins only when $T_2 < T^*$. Taken together these data provide direct confirmation of the appearance of a witness spin glass state at $T^*$ in $ZnCu_3(OH)_6Cl_2$.

This begs the question of whether and how such witness spin interactions could be induced by a QSL. To address this, we consider a conventional Hamiltonian for mutual witness spin interactions

$$H = \sum_{ij} J_{ij} \boldsymbol{s}_i \cdot \boldsymbol{s}_j \tag{1}$$

where $\boldsymbol{s}_i$ is the spin-1/2 witness at lattice position $i$, the sum is over all pairs of such spins in the material, and $J_{ij}$ represents the exchange energy scale between witness spins separated by $r_{ij}$. We here consider a simple model of spinon-mediated interactions through a kagome $Z_2$ QSL[36-40], motivated in part by several earlier studies[41,42,43]. $J_{ij}$ may be estimated by modeling each witness spin as separately coupled to the three closest kagome spins in each of its two neighboring planes (pyramidal links Fig. 1b). In terms of the spin-spin susceptibility of the QSL linking the kagome spin sites $l$,$m$ in the same plane, the witness spin interactions are

$$J_{ij} = \frac{\gamma^2}{4} \sum_{l \in \therefore_i} \sum_{m \in \therefore_j} \zeta_{lm} \tag{2}$$

where $\gamma$ is the strength of the witness spin to kagome coupling, $\therefore_i$ denotes the three kagome sites nearest to witness spin $i$ (pyramidal links Fig. 1b), and $\zeta_{lm}$ is the static spin susceptibility between kagome sites $l$ and $m$ mediated by the QSL. To estimate $\zeta_{lm}$ we evaluate the spinon band structure using a Schwinger fermion mean-field decoupling of the kagome Heisenberg antiferromagnet as detailed in ref. 37. While not as precise as DMRG or exact diagonalization, the model allows us to reach the large system sizes necessary for long-range interactions. For our QSL Hamiltonian, $H_{\mathrm{QSL}}$, the parameters are the nearest neighbor spinon hopping rate $t_1/\hbar$ the second neighbor rate $t_2/\hbar$, the gap parameter, and two Lagrange multipliers which enforce the physical Hilbert space constraint of half-filling. We determine all parameters self-consistently for the kagome plane of herbertsmithite, yielding a gap of $2\Delta \approx 0.44 t_1 \approx 30$ K consistent with previous theoretical studies[36,38,40] (Methods C) and the experimental kagome-lattice susceptibility $\chi_{\mathrm{K}}(T)$ collapsing for $T \lesssim 30$ K [16]. Linear response theory for the QSL gapped parabolic spinon band structure yields an intra-kagome spin susceptibility of

$$\zeta_{lm} = -\frac{2}{\pi} \int_{-\infty}^{E_{\mathrm{F}}} \Im[G_{lm}(E) G_{ml}(E)] \mathrm{d}E\,, \tag{3}$$

where

$$G_{lm}(E) = \langle l | \left( E + i\eta - H_{\mathrm{QSL}} \right)^{-1} \left| m \right\rangle \tag{4}$$

is the real-space spinon Green's function connecting sites $l$ and $m$ at energy $E$, while $\eta$ is an infinitesimal positive regularization. We evaluate $\zeta_{lm}$ using exact diagonalization on clusters of up to 20×40 kagome unit cells. We find good agreement with approximate analytic results obtained for the similar case of electron-mediated interactions in gapped graphene[42,43] (where the parabolic electron bands are due to spin-orbit interactions) in the long-distance limit:

$$J_{ij}(r) \propto \frac{\exp\left(-\frac{r}{r_0}\right)}{r^{\frac{3}{2}}} \tag{5}$$

where $r$ is the separation of witness spins $i$ and $j$, and for the $Z_2$ QSL we would expect

$$r_0 = \sqrt{2} d \left( \frac{t_1}{2\Delta} \right), \tag{6}$$

where $d$ is the nearest-neighbor kagome Cu spacing. Crucially, when spinon-mediated, we find all witness spin interactions are purely antiferromagnetic (AF). This is consistent with the antiferromagnetic nearest-neighbour interaction associated with the spin structure factor from inelastic neutron scattering studies[17] while at the same time generating further neighbor interactions that frustrate the witness spin interactions. Overall, this reveals how the gapped spinon spectrum of a $Z_2$ QSL in herbertsmithite could robustly mediate interactions between quantum witness spins.

One free parameter, $\gamma$ , controls the intensity of spinon mediation of witness spin dynamics. We constrained $\gamma$ by the accepted Curie-Weiss temperature $\theta_{\mathrm{CW}} = -1.1$ K (Methods C), yielding $|\gamma| = 60\ \mathrm{K} \approx \frac{J_{\mathrm{K}}}{3}$ . We note that both ferromagnetic and antiferromagnetic coupling $\gamma = \pm 60$ K are consistent with our model described by $|\gamma|^2$. Further, the order of magnitude of $|\gamma|$ seems plausible because the witness spins and the kagome spins are both $Cu^{2+}$ $s = 1/2$ ions with similar Cu-Cu separations (although their bond angles with the oxygen atoms do differ and so $|\gamma|$ could have been much smaller)[44]. However, this brings into sharp focus a key mystery in herbertsmithite studies: virtually all low-temperature spin dynamical phenomena[14,15,17,26-28,35] occur approximately two orders of magnitude lower in temperature than $J_{\mathrm{K}}$. Most obviously, why is $\theta_{\mathrm{CW}}$ so strongly suppressed below the natural $Cu^{2+}$ $s = 1/2$ interaction scale $J_{\mathrm{K}}$? Spinon-mediation provides a simple and quantitative explanation: nearest neighbor witness spin interactions communicate via spinons propagating through the QSL at third order. The estimated intra-kagome spin susceptibility in dimensionless form, $\zeta'_{lm} \approx t_1\zeta_{lm}$, then reveals from Eqn. (2) that the spinon-mediated witness spin interaction energy scale is

$$\frac{\gamma^2}{4}\frac{\zeta'_{lm}}{t_1} \approx \frac{(60\ \mathrm{K})^2}{4}\frac{0.1}{76\ \mathrm{K}} \approx 1.2\ \mathrm{K} \tag{7}$$

because for the herbertsmithite $Z_2$ QSL theory, Eqn. (3) yields $\zeta'_{lm} \approx 0.1$ for nearest neighbors, and $t_1 = 0.4J_{\mathrm{K}} \approx 76$ K [29]. This agrees strikingly with the experimental value $|\theta_{\mathrm{CW}}| \approx 1.1$ K. Thus, the energy scale of a wide variety of $T \to 0$ spin dynamical characteristics[14,15,17,26-28,35] of herbertsmithite emerges naturally and quantitatively from spinon mediation of witness spin dynamics.

To understand the global consequences requires large-scale simulations of spinon-mediated witness spin dynamics. Accordingly, we simplified Eqn. (1) by approximating the quantum spins with classical Ising variables, enabling us to conduct Monte Carlo (Metropolis–Hastings) simulations. The validity of this approximation for use in herbertsmithite witness spin dynamics simulations is discussed in detail in Methods D. Each witness spin interacts with all others that share a kagome plane. We use our numerically calculated $J_{ij}(r)$, Eqn. (2), with 45×45×4 possible witness spin sites (randomly occupied to ~33% per $Zn^{2+}$ plane) and periodic boundary conditions (Methods D). We present the interaction energy $J_{ij}(r)$ between a single witness spin at the centre and one at any other witness spin site, along with the analytic approximation, in Fig. 4a. An important consequence stemming from $J_{ij}(r)$ is that, while the known nearest neighbor AF witness spin interactions would lead to AF order (above the percolation threshold), the extensive spinon-mediated interactions frustrate this ordering. This generates a plethora of physical effects. Most elementary is the time-sequence of predicted magnetization fluctuations $M(t,T) \propto \sum_i s_i(t,T)$ where $i$ represents all witness spin sites in the simulated crystal. The fluctuations in witness spin $M(t,T)$ are predicted to intensify and slow as falling $k_{\mathrm{B}}T$ approaches $J_{ij}(r = d)s_i s_j$, as shown in Fig. 4b. The fluctuations have Gaussian distributions (Extended Data Fig. 4a). The predicted power spectral density of this witness spin magnetization noise $S_M(\omega, T)$ (measured in units of radians/MCS where MCS is the Monte-Carlo time step) is shown in Fig. 4b inset and Fig. 4c, while Fig. 4d shows the contour plot of $S_M(\omega, T)$ predictions revealing a transition in witness spin dynamics at $T^* \approx 150$ mK (horizontal arrow). Further, the witness spin noise power-index $\alpha(T)$ derived from fitting $S_M(\omega, T) \propto \omega^{-\alpha(T)}$ to the data in Figs. 4c,d transitions to a constant $\alpha \approx 1$ at $T^*$ (Fig. 5a) (Methods D). The witness spin noise variance $\sigma_M^2(T)$ from Figs. 4c,d exhibits a transition in noise power at $T^*$ below which it collapses (Fig. 5b). Moreover, although the DC magnetic susceptibility of the witness spins approximately follows a Curie-Weiss trajectory at higher temperatures near $T \approx 1$ K (Fig. 5c), once the spinon-mediated interactions become predominant, the $Z_2$ QSL model predicts a sharp cusp in $\chi(T)$ at a transition temperature near 150 mK, as shown in Fig. 5c. Finally, analysis using all predicted witness spin configurations $s_i$ in terms of either an AF or EA spin-glass order parameter (Fig. 5d), reveals the finite EA (and zero AF) order parameter indicating witness spin freezing. By juxtaposing the diverse predictions from Eqns. (1-4) for spinon-mediated witness spin dynamics and the ground state as presented in Fig. 5, with the discovered witness spin dynamical phenomenology of herbertsmithite as illustrated in Fig. 3, their conspicuous correspondence evidently validates the spinon-mediated witness spin dynamics concept (see also Extended Data Fig. 5).

This revelation resolves open mysteries of the $T \to 0$ phenomenology of herbertsmithite. While $Cu^{2+}$/$Zn^{2+}$ substitution has long been adduced as the trigger for such effects[6-8,11,12-21,23-28,35,45-54], no specific microscopic theory has previously been put forward. Indeed, the opposite approach has been more typical, with much effort expended in minimizing their influence upon results. Here we explore the obverse perspective: identification of a specific microscopic mechanism and its quantitative theory for witness spin interactions via the QSL. Thence, we find the susceptibility $\chi^{-1}(T) \propto (T - \theta_{\mathrm{CW}})$ with $\theta_{\mathrm{CW}} \approx -1.1$ K is highly consistent with the spinon-mediated $J_{ij}(r)$ of the witness spin interactions (Fig. 5c). Similarly, we confirm (Methods E; Extended Data Fig. 6) that the low-energy neutron scattering structure factor $\Sigma(\boldsymbol{q}, \omega < 1\ \mathrm{meV})$ of herbertsmithite[17,27] is not inconsistent with witness spin dynamics controlled by the spinon-mediated $J_{ij}(r)$ between pairs of nearest-neighbor witness spins (Fig. 4a). With respect to $^{17}O$ nuclear magnetic resonance relaxation rates, $1/T_1(T)$, their sharp diminution below $T \approx 250 \pm 50$ mK [48] is consistent with the suppression of witness spin fluctuations below $T^*$, as predicted within our spinon-mediation theory (Fig. 5b).

Our empirical knowledge of herbertsmithite witness spin interactions with its QSL has thus been augmented and clarified by the introduction of spin noise spectroscopy[30-34] to QSL studies. This reveals the existence (Figs. 2a,b), slowing and intensification of scale-invariant witness spin noise with power spectral density $S(\omega, T) \propto \omega^{-\alpha(T)}$ (Figs. 2b,c,d). These fluctuations evolve to reach a constant $\alpha \approx 1$ at $T^* \approx 260$ mK (Fig. 3a), at which point the spin noise power $\sigma_M^2(T)$ begins to diminish steeply (Fig. 3b). Moreover at this $T^*$, the susceptibility $\chi(T)$ (measured at previously unexplored μT fields) experiences a sharp transition into a witness spin glass phase (Fig. 3c), exhibiting an EA spin-glass order-parameter and ultra-slow relaxation (Fig. 3d). All these phenomena are consistent within a model having a $2\Delta \approx 0.44 t_1 \approx 30$ K gapped spinon spectrum of a $Z_2$ QSL in herbertsmithite (Methods C), thus explaining the $T \to 0$ phenomenology if the coupling constant between $Cu^{2+}$ witness spins and kagome spins is $|\gamma| \approx J_\mathrm{K}/3$. The form and range of the witness spin interaction function $J_{ij}(r)$ (Fig. 4a) and, more fundamentally, the quantitative form and structure of the spin-spin susceptibility within the $Z_2$ QSL, are then determined theoretically.

We have considered a range of alternative models, including random spin singlets, Zn substitution into the kagome planes, witness spin dipole-dipole interactions, spin-wave mediated witness spin interactions, ferromagnetic clusters, and local magnetic exchange mechanisms and all can be ruled out based on the experimentally determined phenomenology as $T \to 0$ as explained in full detail in Methods F and Extended Data Fig. 7. Only spinon mediation shows consistency with the full range of experimental data (Fig. 3), and that of a $Z_2$ QSL (Fig. 5) gave a somewhat better quantitative match than a U(1) QSL (Extended Data Fig. 8). However, it remains challenging to give a definitive discrimination between $Z_2$ and U(1) spinon-mediation of witness spin interactions. A comprehensive fully quantum mechanical theory for the influences of witness spins on the kagome plane and vice versa may be required. This could greatly improve the precision of the theoretical model, for example, the spin noise spectral shape above $T^*$ and the detailed spin dynamics as $T \to 0$ (Methods D, E). Such a study might also quantify the extent to which witness spins could affect the quantum state in the kagome plane, further enhancing the efficacy of the witness spin approach to probing quantum spin liquids.

Notwithstanding these theoretical challenges, further exceptional opportunities for QSL studies also emerge. First, $Cu^{2+}$/$Zn^{2+}$ substitutional atoms no longer obscure the QSL physics of herbertsmithite, but, instead, they would potentially allow direct quantum detection and interrogation of the spinon spectrum (Figs. 3,5). Second, the herbertsmithite witness spin glass state as $T \to 0$ appears to be a unique state of quantum matter, predicated upon long-range quantum entanglement through a QSL. Third, witness spins may now be used as a 'quantum portal' through which to access, manipulate and transit the QSL. Finally, these witness spin noise techniques and interpretations are eminently generalizable to wide-ranging QSL research in other target materials e.g. Zn-barlowite, as well as to similarly elusive states such as random-bond quantum dimer systems.

**Acknowledgments:** We acknowledge and thank C. Carroll, J.T. Chalker, R. Coldea, J.Dasini, P.J. Hirschfeld, P.A. Lee, S.A. Kivelson, G. Luke, P. Mendels, S.L. Sondhi, F. Turci, N. Wilding, and A. P. Young for discussions and guidance. S.J.B. acknowledges support from UK Research and Innovation (UKRI) under the UK government's Horizon Europe funding guarantee [Grant No. EP/X025861/1]. F.F. acknowledges support from EPSRC Grant No. EP/X012239/1. M.R.N. was supported by the Materials Sciences and Engineering Division, Basic Energy Sciences, Office of Science, U.S. Dept. of Energy. C.-C. H. and J.C.S.D. acknowledge support from the European Research Council (ERC) under Award DLV-788932. J.C.S.D. acknowledges support from the Moore Foundation's EPiQS Initiative through Grant GBMF9457. H. Takahashi and J.C.S.D. acknowledge support from the Royal Society under Award R64897. J.M., J.W., and J.C.S.D. acknowledge support from Science Foundation of Ireland under Award SFI 17/RP/5445.

**Author Contributions Statement:** H. Takagi, F.F. and J.C.S.D. conceived the project. P.P., M.I., Y.M., and H. Takagi synthesized and characterized the samples; J.M. C.-C.H, F.J., J.W., and H. Takahashi developed the millikelvin spin noise spectrometer and carried out experimental measurements; M.R.N. provided theoretical guidance; H. Takahashi and M.R.N. performed data analysis; M.W.-T., M.A.S.-M. and F.F. developed the spinon-mediated witness spin dynamics theory and executed simulations in collaboration with S.J.B., and M.W.-T. carried out simulations. F.F., M.R.N. and J.C.S.D. supervised the research and wrote the paper with key contributions from M.W.-T. and H. Takahashi. The manuscript reflects the contributions and ideas of all authors.

**Competing Interests Statement:** The authors declare no competing interests.

## Figure Legends

### Fig. 1 Witness spins in $ZnCu_3(OH)_6Cl_2$

**a.** Crystal structure of $ZnCu_3(OH)_6Cl_2$ viewed along the *c*-axis showing the kagome plane of $Cu^{2+}$ $s = 1/2$ ions hypothesized to sustain a QSL, and indicating sites of $Cu^{2+}$ $s = 1/2$ ions substituted onto non-magnetic $Zn^{2+}$ $s = 0$ sites.

**b.** Interplanar perspective of $ZnCu_3(OH)_6Cl_2$ again indicating sites of $Cu^{2+}$ $s = 1/2$ ions acting as witness spins substituted on 33% of $Zn^{2+}$ sites. Coupling of each witness spin to three $Cu^{2+}$ spins in the adjacent kagome plane is indicated by pyramidal links.

**c.** The schematic design of the SQUID spin noise spectrometer that measures the time evolution of flux $\Phi(t)$ generated by the $ZnCu_3(OH)_6Cl_2$ sample in the superconductive pickup coil connected persistently to the SQUID input coil.

### Fig. 2 Spin noise spectroscopy of $ZnCu_3(OH)_6Cl_2$

**a.** Typical time sequence of the measured magnetic flux $\Phi(t)$ generated by $ZnCu_3(OH)_6Cl_2$ at 260 mK (light yellow dots), where frequency components out of bandwidth 0.3 rad•Hz $\leq \omega \leq$ 600 rad•Hz are filtered out. For visual clarity, the plotted data points are down-sampled to every 5 ms. The box-car average of the signal for every 50 ms is overlaid (dark yellow), which is highly distinct from the identically averaged signal of the empty coil (gray).

**b.** Typical examples of the squared flux $\Phi^2(t,T)$ at eight selected temperatures. Again, the signal bandwidth is 0.3 rad•Hz $\leq \omega \leq$ 600 rad•Hz and points are down-sampled to every 0.5 ms. Inset: power spectral density $S_\Phi(\omega,T)$ at four selected temperatures after subtracting background contributions. Error bars are the standard error of segment averaging (Methods B).

**c.** Power spectral density $S_\Phi(\omega,T)$ of measured $ZnCu_3(OH)_6Cl_2$ witness spin flux noise, which spans a frequency range of at least 0.6 rad•Hz $\leq \omega \leq$ 600 rad•Hz. The spectra exhibit scale-invariant forms $\omega^{-\alpha}$. The equivalent power spectral density $S_M(\omega,T)$ of magnetization noise at the sample is presented in units of Tesla on the right hand axis.

**d.** Contour plot of measured $S_\Phi(\omega,T)$ from Fig. 2c revealing a clear transition in witness spin dynamics at $T^* \approx 260\ \mathrm{mK}$ (horizontal arrow).

**Fig. 3 Spin noise spectroscopy analysis of witness spin dynamics in $ZnCu_3(OH)_6Cl_2$**

**a.** Measured witness spin flux noise power-index $\alpha$ from $S_\Phi(\omega,T) \propto \omega^{-\alpha(T)}$ as a function of temperature from Fig. 2c, obtained by fitting the power spectral density $S_\Phi(\omega,T) = A(T)\omega^{-\alpha(T)}$ in the frequency range 0.6 rad•Hz $\leq \omega \leq$ 600 rad•Hz. A clear transition to $\alpha \approx 1$ is detected at $T^* \approx 260\ \mathrm{mK}$ (dashed line).

**b.** Measured witness spin flux noise variance $\sigma_\Phi^2$ as a function of temperature from Fig. 2c, obtained by integrating $S_\Phi(\omega,T)$ in the range $0.6\ \mathrm{rad{\bullet}Hz} \leq \omega \leq 600\ \mathrm{rad{\bullet}Hz}$. Right axis: the equivalent magnetization noise variance $\sigma_M^2(T)$. These data indicate a transition in noise power at $T^* \approx 260\ \mathrm{mK}$ (dashed line).

**c.** Measured witness spin susceptibility $\chi(T)$ (SI units, inset shows $1/\chi$) in micro-Tesla magnetic fields revealing a Curie-Weiss-like $\chi(T)$ at higher temperatures (orange line), yielding an estimated witness spin density of 33% of Zn sites. However, this diverging $\chi(T)$ is interrupted by a sharp transition to a rapidly diminishing $\chi(T)$ below $T^* \approx 260\ \mathrm{mK}$ (dashed line).

**d.** Measured Edwards-Anderson spin-glass order parameter for witness spins from Fig. 3c, indicating a transition to a witness spin glass below $T^* \approx 260\ \mathrm{mK}$ (dashed line). Inset: time evolution of the average flux against a 2 $\mu\mathrm{T}$ applied field after the temperature is suddenly dropped from a thermalized condition at 400 mK. The sample flux $\Phi(t)$ shows a $-\ln t$ relaxation (dashed line) on periods of a day only below $T \approx 250\ \mathrm{mK}$, providing direct evidence of the appearance of a witness spin glass.

Error bars are invisible as they are smaller than the symbol points in all panels.

**Fig. 4 Simulated spinon-mediated witness spin noise dynamics in $ZnCu_3(OH)_6Cl_2$**

**a.** Spinon-mediated interaction energy $J_{ij}(r)$ between witness spins. Left: numerically calculated interactions due to a single $Cu^{2+}$ witness spin at the origin, Eqn. (2). The gray color indicates $J_{ij}(r) < 0$. Right: analytic long-distance approximation, Eqn. (5).

**b.** Typical example of the simulated magnetization dynamics $M(t)$ of witness spins under extensive spinon-mediated exchange interactions, plotted as $M^2(t,T)$ at eight selected temperatures. Frequency components out of bandwidth $3 \times 10^{-5}$ rad/MCS $\leq \omega \leq 6 \times 10^{-2}$ rad/MCS are filtered out, and points are down-sampled to every 5

MCS. Inset: power spectral density $S_M(\omega, T)$ at four selected temperatures. Error bars are the standard error of segment averaging (Methods D).

**c.** Predicted power spectral density $S_M(\omega, T)$ of witness spin magnetization noise as a function of frequency and temperature due to spinon-mediated interactions.

**d.** Contour plot of predicted $S_M(\omega, T)$ from Fig. 4c revealing a clear transition in dynamics at $T^* \approx 150\ \mathrm{mK}$ (horizontal arrow).

**Fig. 5 Spectroscopy analysis of simulated spinon-mediated witness spin dynamics in $ZnCu_3(OH)_6Cl_2$**

**a.** Predicted witness spin magnetization noise power-index $\alpha$ for $S_M(\omega, T) \propto \omega^{-\alpha(T)}$ as a function of temperature from Fig. 4c, revealing a transition to $\alpha \approx 1$ at $T^*$ (dashed line). The open circles are used for temperatures above $T^*$ where power-law fitting is challenging. Error bars are the standard error from fitting.

**b.** Predicted witness spin magnetization noise variance $\sigma_M^2$ as a function of temperature from Fig. 4c, indicating a transition in noise power at $T^*$ (dashed line).

**c.** Predicted witness spin-only susceptibility $\chi(T)$ (SI units, inset shows $1/\chi$) due to spinon-mediated interactions, revealing a sharp transition at $T^*$ (dashed line) from a Curie-Weiss behaviour determined at higher temperatures (orange line).

**d.** Predicted antiferromagnetic order parameter (green crosses) and Edwards-Anderson spin-glass order parameter (black dots) from witness spin simulations, indicating that $T^*$ (dashed line) is the transition to a spinon-mediated witness spin glass.

Error bars are invisible as they are smaller than the symbol points in **b**,**c**,**d**.

# Figure 1

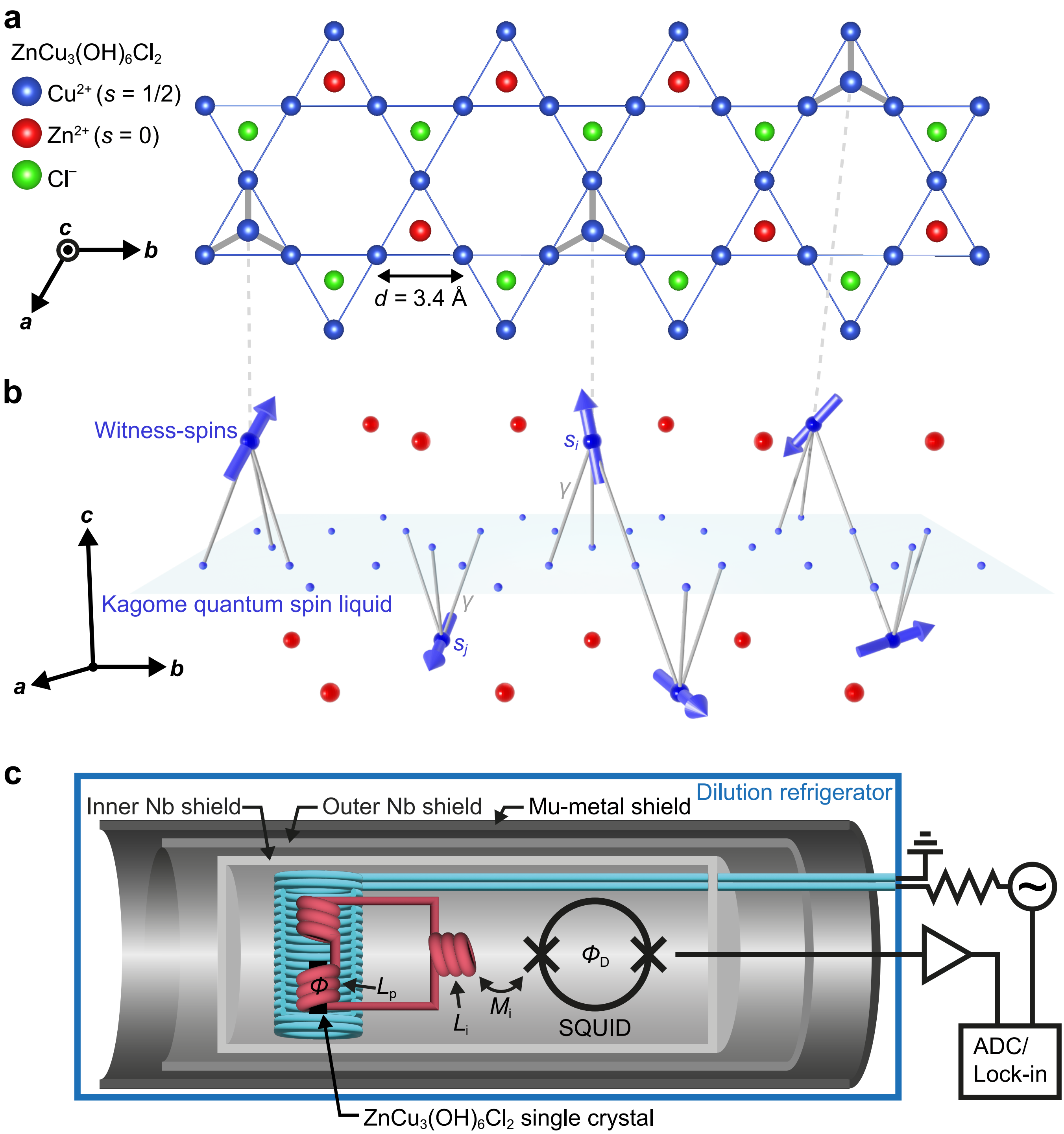

**a**

$ZnCu_3(OH)_6Cl_2$

$Cu^{2+}$ ($s$ = 1/2)

$Zn^{2+}$ ($s$ = 0)

$Cl^-$

*c*

*b*

*a*

$d$ = 3.4 Å

**b**

Witness-spins

$s_i$

$\gamma$

*c*

Kagome quantum spin liquid

$\gamma$

$s_j$

*a*

*b*

**c**

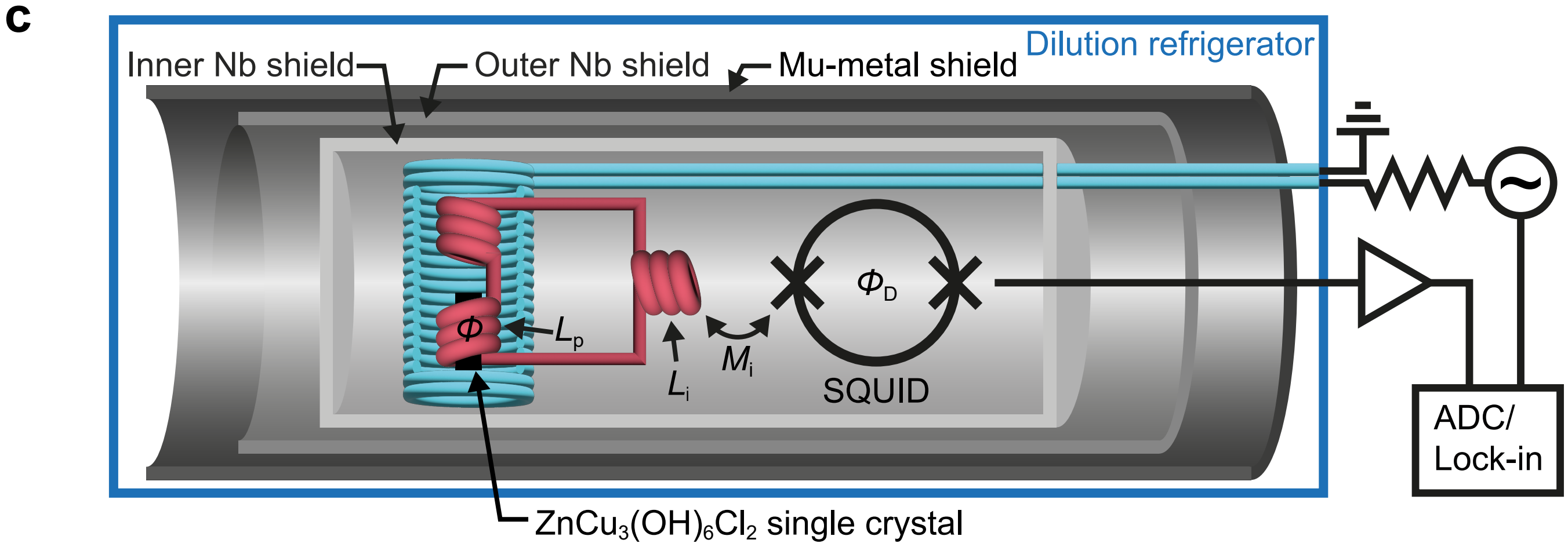

# Figure 2

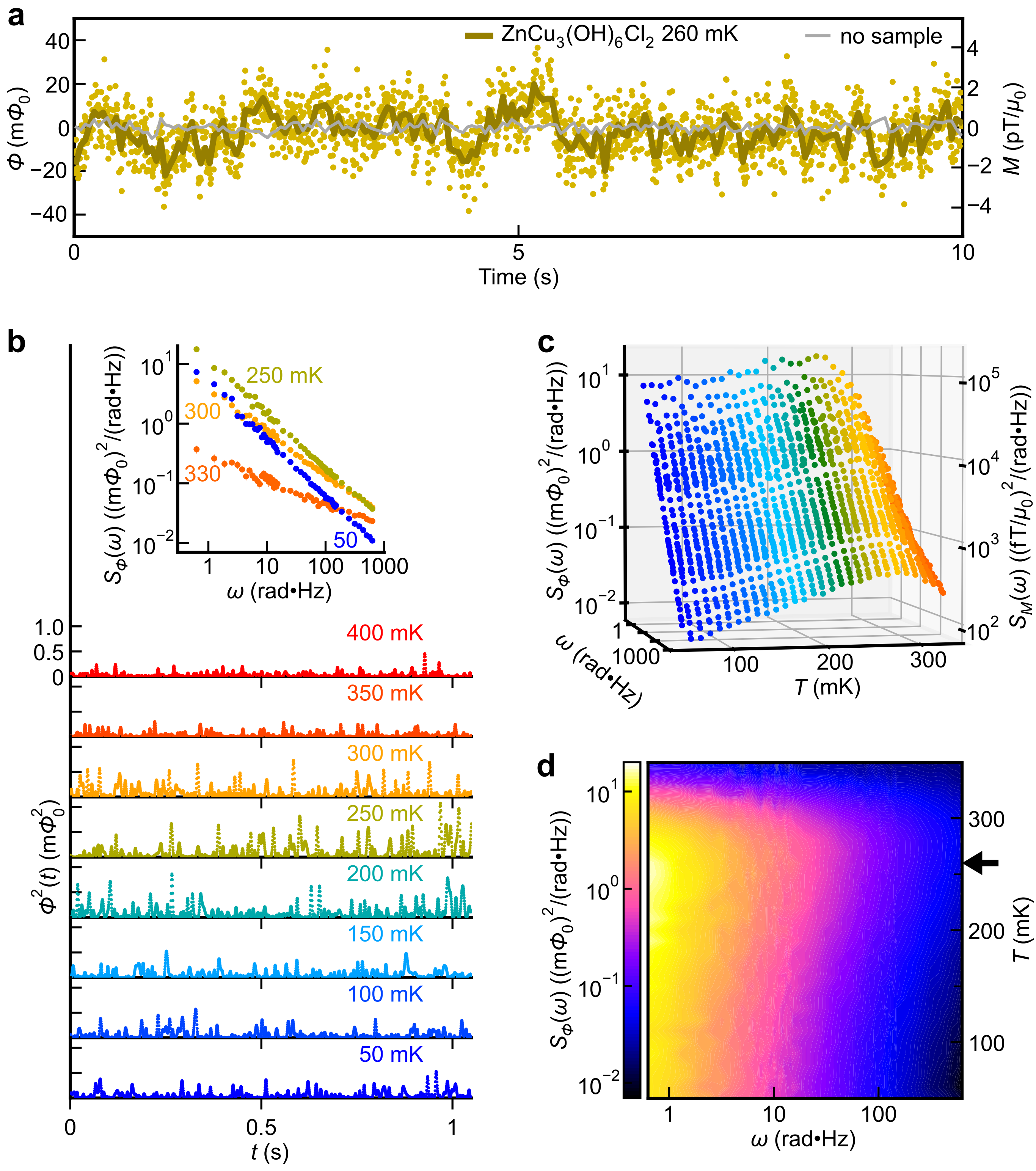

# Figure 3

# Figure 4

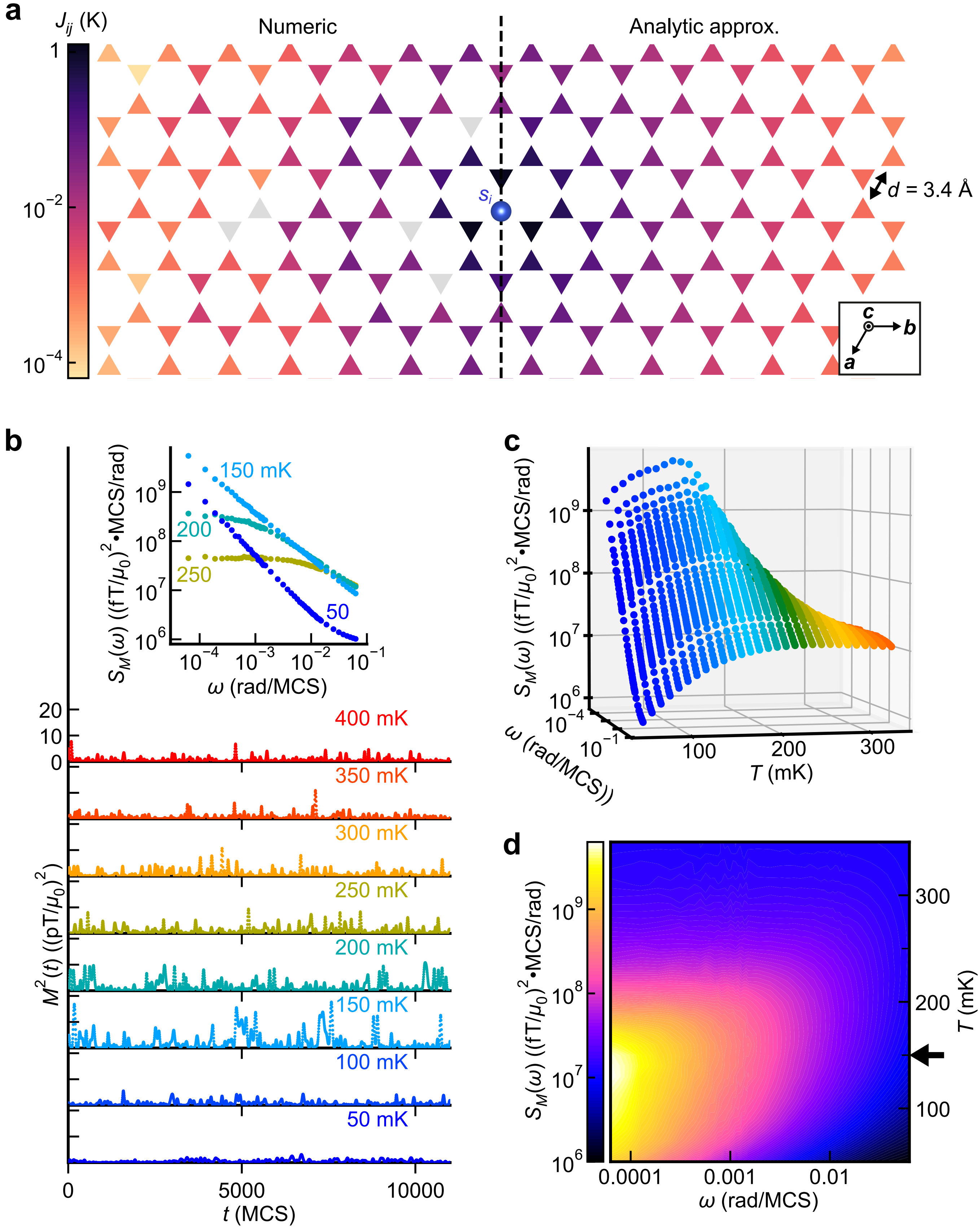

# Figure 5

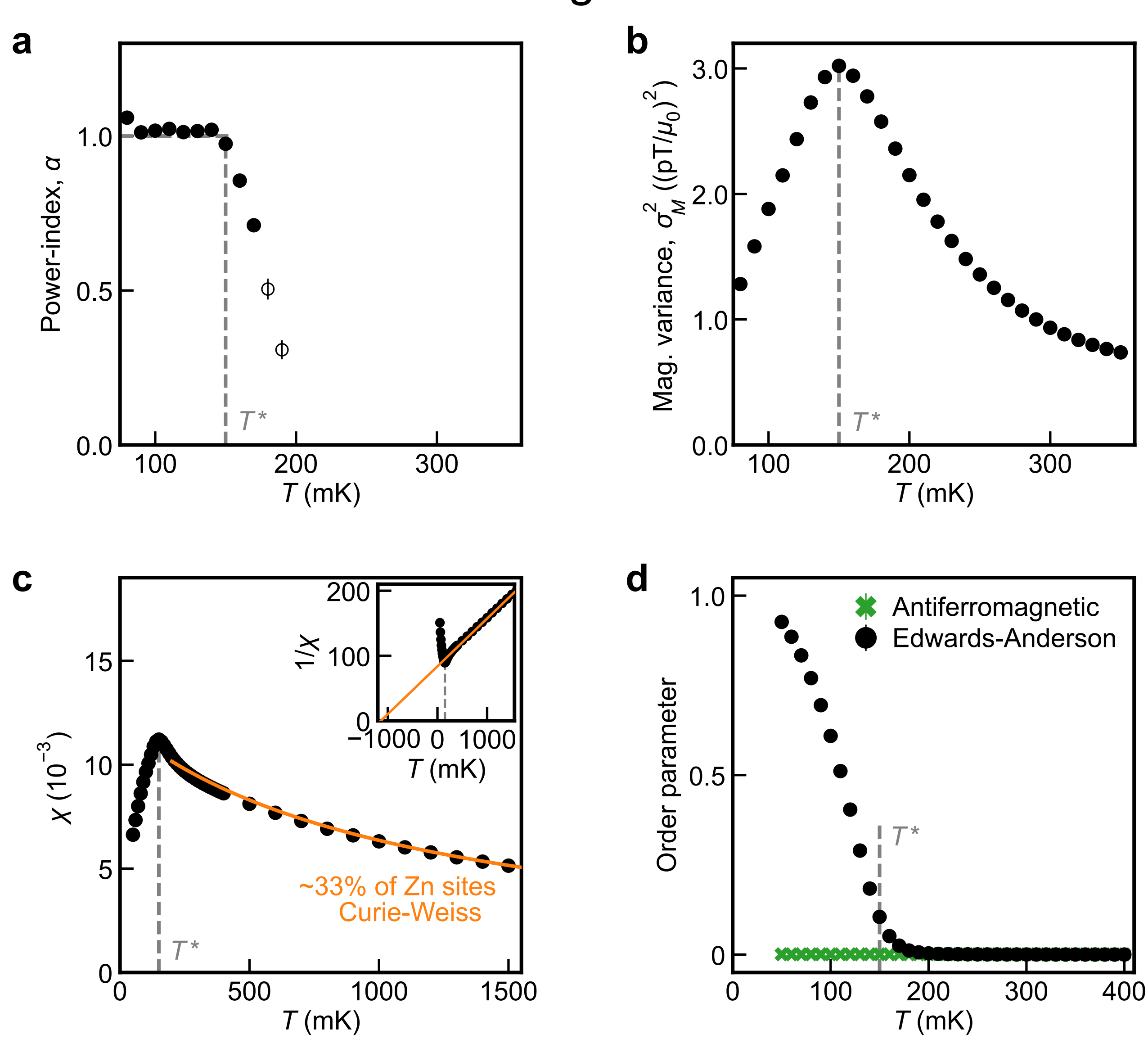

## Methods:

### A. Herbertsmithite samples

$ZnCu_3(OH)_6Cl_2$ single crystals were synthesized as described in ref. 55, using a recrystallization method. Powders of $ZnCl_2$, CuO, and $H_2O$ were mixed in a quartz tube with a ratio of 2.015 g : 0.235 g : 4.5 mL. The tube was sealed under vacuum and laid horizontally in a three-zone gradient furnace, with the temperature of hot and cold ends set at 180 °C and 160 °C, respectively. Millimeter-scale single crystals were obtained after 3 months. Extended Data Fig. 1a shows the photos of the three $ZnCu_3(OH)_6Cl_2$ single crystals studied. The lattice structure was confirmed by x-ray Laue diffraction, as exemplified by the clear Bragg peaks of Sample 1′ shown in Extended Data Fig. 1b. Measurements were performed along the $c$-axis for Samples 1 and 2 and the $a$-axis for Sample 3.

The stoichiometry of Zn:Cu ratio is found to be 0.97:3.03 for the samples reported here, by using inductively coupled plasma mass spectrometry. Refinement of a single crystal x-ray diffraction measurement indicates that 32.5% of the $Zn^{2+}$ sites and 10.8% of the $Cu^{2+}$ sites are inter-substituted[25]. Extended Data Fig. 1c shows the DC susceptibility (Sample 1') in SI units measured by a SQUID MPMS (Quantum Design) and its Curie-Weiss fitting by $\chi = \chi_0 + \frac{C_{\text{Curie}}}{T-\theta_{\text{CW}}}$. Fitting in a temperature range 150 K $\leq T \leq$ 320 K yields $\chi_0 = -5 \times 10^{-6}$, $\theta_{\text{CW}} = -280$ K, and $C_{\text{Curie}} = 0.165$ K. Fitting in a temperature range 2 K $\leq T \leq$ 6 K yields $\chi_0 = (4.3 \pm 0.2) \times 10^{-4}$ , $\theta_{\text{CW}} = -1.07 \pm 0.03$ K, and $C_{\text{Curie}} = 0.0134 \pm 0.0002$ K corresponding independently to $S$ = 1/2 at 32.5±0.5% of the Zn sites. The Curie-Weiss fitting of the low-temperature DC susceptibility is stable as long as the fitting range is within 2 K $\leq T \leq$ 10 K, although it becomes sharply fitting-range-dependent below 1 K where the susceptibility starts diverting from the Curie-Weiss behavior (Fig. 3c)[14]. These sample characterization results are comparable to the past stoichiometry studies[11,20,45], neutron diffraction studies[11,20,56], and DC susceptibility studies[14,35,57].

As the Cu occupation probability of witness spin sites in our single crystals, we take 33%. This value is based on the coincidence of estimates from the single crystal x-ray diffraction measurement[25] and DC susceptibility Curie-Weiss fitting (Extended Data Fig. 1c), both of which are performed on our single crystal (Sample 1'). We note that the precise nature of site-disorder in herbertsmithite has not yet been fully determined[11,23,24,58], with reported Cu substitution percentage at Zn sites ranging from 12% [17] to 36% [56] , and of Zn substitution percentage at Cu ranging from 0% [11,58] to 10% [23,56].

### B. Measurements

Magnetic flux $\Phi(t)$ noise measurements were performed with a 19-turn single superconducting pickup coil connected to a DC SQUID SQ1200 (Star Cryoelectronics), which is designed to maximize the noise measurement sensitivity. Susceptibility measurements were performed with a 10-turn-each in-series counter-wound superconducting pickup coil connected to a SQUID SP550 (Quantum Design) (Fig. 1c), using a solenoid whose magnetic field was calibrated by using an indium cylinder in a superconducting state ($\chi = -1$). In both setups, three 0.2-mm-diameter silver wires were directly attached to the sample for thermalization, and measurements were taken at least 20 minutes after the target temperature was reached. Both setups were shielded by multiple nested niobium and mu-metal cylinders.

Flux noise data at each temperature was recorded for 1000 s at 20 kSa/s For susceptibility measurements at each temperature, the sample magnetic response was recorded as the magnetic field $\mu_0 H$ was swept over 0 μT → $-4$ μT → 4 μT → $-4$ μT → 0 μT in steps of ~ 0.05 μT (zero-field cooling). The slope of this $(\mu_0 M)/(\mu_0 H)$ data, representing the magnetic susceptibility, is extracted by a linear fitting with standard error bars from the linear fit. The micro-Tesla DC susceptibility of Sample 1 in Fig. 3c is obtained by subtracting an offset constant. This offset value is determined so that the measured micro-Tesla susceptibility smoothly connects to the measurement result in the MPMS in the overlapping temperature range 2 K $\leq T \leq$ 3 K (Extended Data Fig. 1d). For the long-term spin evolution under a 2 μT field, a sample was thermalized at $T_1 = 400$ mK for 1 hour and the temperature was rapidly dropped to a lower temperature $T_2$ in less than 5 minutes. After 20 mins of thermalization at $T_2$ (i.e. starting from $t = 1200$ s), the spin evolution was recorded for 80000 s (~1 day) at 1 kSa/s. In Fig. 3d inset, the signal is averaged for every 100 s.

Flux noise data are processed in a similar method as ref. 33. The distribution of $\Phi(t)$ is Gaussian with the expected statistical fluctuations (Extended Data Fig. 2a). The power spectral density (PSD) with frequency resolution $\Delta\omega = (2\pi \text{ rad}) \times (0.1 \text{ Hz}) = 0.6$ rad•Hz is first calculated from 100 split segments[33], with its error bars determined by the standard error of segment averaging. The empty-coil measurement result is subtracted as a background contribution. This PSD is plotted in Figs. 2c,d and Extended Data Fig. 2b. To increase the signal-to-noise ratio, the PSD is averaged over a $10\Delta\omega$ or $100\Delta\omega$ window at high frequencies. The power-law index in Fig. 3a is obtained by fitting the $\Delta\omega = 0.6$ rad•Hz PSD with $S_\Phi(\omega, T) \propto \omega^{-\alpha(T)}$ in the range 0.6 rad•Hz $\leq \omega \leq$ 600 rad•Hz, as shown in Extended Data Fig. 2b. Error bars are the standard error from fitting. The variance $\sigma_\Phi^2$ in Fig. 3b is calculated by integrating the $\Delta\omega = 0.6$ rad•Hz PSD in the range 0.6 rad•Hz $\leq \omega \leq$ 600 rad•Hz (error bars are propagated from the PSD). This is demonstrably equivalent to the variance directly calculated from the time sequence of $\Phi(t)$ (appropriately filtered to the corresponding frequency range), and the variance peak temperature remains close to $T^*$ for all different frequency-integration ranges (Extended Data Fig. 2c).

The Edwards-Anderson spin-glass order parameter $q_{\mathrm{EA}}$ in Fig. 3d is extracted from the micro-Tesla DC susceptibility by solving the following formula[59,60]

$$\chi(T) = \frac{C(1 - q_{\mathrm{EA}}(T))}{T - \theta(1 - q_{\mathrm{EA}}(T))} \;. \tag{8}$$

The constants $C$ and $\theta$ are obtained by fitting in the temperature range above $T^*$ where $q_{\mathrm{EA}}(T)$ vanishes: 270 mK $\leq T \leq$ 500 mK. Error bars of $q_{\mathrm{EA}}(T)$ propagate from $\chi(T)$ and the standard error of the fitted parameters $C$ and $\theta$.

The spin noise data in Figs. 2, 3 were measured in Sample 1. The equivalent measurements were performed for Sample 2 and Sample 3. As shown in Extended Data Fig. 3, the transition at $T^* \approx 260$ mK in the witness spin noise power-index $\alpha$ from the power spectral density $S_\Phi(\omega, T) \propto \omega^{-\alpha(T)}$, the witness spin noise variance $\sigma_\Phi^2$, and the micro-Tesla susceptibility $\chi$; and the $-\ln t$ relaxation of the sample flux $\Phi(t)$ below $T^*$ are reproduced in multiple samples. The Sample 3 response was smaller compared to the other two samples because the small crystal size made it difficult to fill the full length of the pickup coil. Accordingly, the PSD fitting range is limited to 0.6 rad•Hz $\leq \omega \leq$ 60 rad•Hz and the susceptibility measurement result in Extended Data Fig. 3c is scaled for comparison with Sample 1.

The witness spin dynamics and associated transition that we observe in all herbertsmithite samples had not been previously measured in either AC or DC susceptibility studies[14,45]. One possible reason could be a difference in measurement conditions.

Pioneering DC susceptibility measurements[14] were performed at magnetic fields near $B$ =0.05 T. While that field is small compared to the energy scale of the observed transition temperature $T^* \approx 260$ mK, it is empirically known that quite a small field can significantly suppress a sharp peak signature of spin glass transition, making it difficult to detect. For example, a $B$ =0.04 T DC field is capable of suppressing the sharp peak signature of a 21.5 K spin glass transition in $Fe_{0.5}Mn_{0.5}TiO_3$ [61], 0.05 T for 17 K transition in $CdCr_{1.7}In_{0.3}S_4$ [62], 0.06 T for 15.5 K transition in $Gd_{0.37}Al_{0.63}$ [63], and 0.04 T for 0.2 K transition in $Gd_3Ga_5O_{12}$ [64]. Another possible reason is the difference between powder samples and single crystals. The herbertsmithite susceptibility studies[14,45] were performed on powder samples before the establishment of single crystal growth[55]. Differences between single crystal and such powder samples might possibly be caused by different chemical compositions, enhanced surface effects, randomized direction of applied field, which may have prevented observation of a spin glass transition therein. All of the DC susceptibility measurements reported in the present work were carried out at $B \leq 5$ μT on single crystals, and all yield a sharp transition at a virtually identical $T^* \approx 260$ mK, supporting the plausible conclusion that this phenomenon is intrinsic to herbertsmithite single crystals in ambient magnetic fields $|B| \leq$ 5 μT.

### C. Spinon-mediated interactions via $Z_2$ quantum spin liquid

The Hamiltonian for mutual witness spin interactions in Eqns. (1,2) [41,65] is derived from the coupling between a witness spin $\boldsymbol{s}_i$ and a kagome-spin $\boldsymbol{s}_l^{\text{Kagome}}$

$$H_{\text{coupling}} = \gamma \boldsymbol{s}_i \cdot \boldsymbol{s}_l^{\text{Kagome}}. \tag{9}$$

The intra-kagome spin susceptibility $\zeta_{lm}$ is calculated using linear response Eqns. (3,4) from the spinon band structure of a $Z_2[0,\pi]\beta$ QSL. A $Z_2[0,\pi]\beta$ QSL is the only gapped QSL that is compatible with lattice symmetries at the mean-field level, and is in the neighborhood of the $U(1)[0,\pi]$ state whose energy is the lowest among different U(1) QSL[29].

In units of the nearest neighbor spinon hopping energy $t_1 = 0.4 J_{\text{K}} \approx 76$ K [29], the $Z_2[0,\pi]\beta$ QSL Hamiltonian contains real parameters for the second neighbor hopping $t_2$, gap $\Delta_2$, and two Lagrange multipliers $\lambda_1$, $\lambda_3$ which enforce the physical Hilbert space constraint of half-filling. We solve for all four using the standard self-consistent mean field approach. ref. 37 gives the following mean field spinon Hamiltonian for the kagome planes:

$$\boldsymbol{s}_l^{\text{Kagome}} = \frac{1}{2} \sum_{\alpha,\beta=\{\uparrow,\downarrow\}} f_{i\alpha}^{\dagger} \sigma_{\alpha\beta} f_{i\beta} \,, \tag{10}$$

$$\begin{aligned} \hat{H}_{\text{QSL}} = & \left\{ \sum_i^N \lambda_3 \left( f_{i\uparrow}^{\dagger} f_{i\uparrow} + f_{i\downarrow}^{\dagger} f_{i\downarrow} \right) + \lambda_1 \left( f_{i\uparrow}^{\dagger} f_{i\downarrow}^{\dagger} + f_{i\downarrow} f_{i\uparrow} \right) \right\} \\ & + \left\{ \sum_{ij} \left( t_1 \nu_{ij}^{(1)} + t_2 \nu_{ij}^{(2)} \right) \left( f_{i\uparrow}^{\dagger} f_{j\uparrow} + f_{i\downarrow}^{\dagger} f_{j\downarrow} - f_{i\uparrow} f_{j\uparrow}^{\dagger} - f_{i\downarrow} f_{j\downarrow}^{\dagger} \right) \right. \\ & \left. + \Delta_2 \nu_{ij}^{(2)} \left( f_{i\uparrow}^{\dagger} f_{j\downarrow}^{\dagger} - f_{i\downarrow}^{\dagger} f_{j\uparrow}^{\dagger} - f_{i\uparrow} f_{j\downarrow} + f_{i\downarrow} f_{j\uparrow} \right) \right\} , \end{aligned} \tag{11}$$

where $\nu_{ij}^{(1)}$ is non-zero only for first nearest neighbors (and is 1 or $-1$ as defined in ref. 37), and $\nu_{ij}^{(2)}$ is non-zero only for second nearest neighbors. There are $N$ sites in the system. It is convenient to rewrite the diagonal terms using

$$\{f_i, f_j^{\dagger}\} = \delta_{ij} \,, \tag{12}$$

$$\{f_i, f_j\} = 0, \tag{13}$$

to give

$$\begin{aligned} \hat{H}_{\text{QSL}} = N\lambda_3 + & \left\{ \sum_i \frac{\lambda_3}{2} \left( f_{i\uparrow}^{\dagger} f_{i\uparrow} - f_{i\uparrow} f_{i\uparrow}^{\dagger} + f_{i\downarrow}^{\dagger} f_{i\downarrow} - f_{i\downarrow} f_{i\downarrow}^{\dagger} \right) + \frac{\lambda_1}{2} \left( f_{i\uparrow}^{\dagger} f_{i\downarrow}^{\dagger} + f_{i\downarrow} f_{i\uparrow} - f_{i\downarrow}^{\dagger} f_{i\uparrow}^{\dagger} - f_{i\uparrow} f_{i\downarrow} \right) \right\} \\ & + \left\{ \sum_{ij} \left( t_1 \nu_{ij}^{(1)} + t_2 \nu_{ij}^{(2)} \right) \left( f_{i\uparrow}^{\dagger} f_{j\uparrow} + f_{i\downarrow}^{\dagger} f_{j\downarrow} - f_{i\uparrow} f_{j\uparrow}^{\dagger} - f_{i\downarrow} f_{j\downarrow}^{\dagger} \right) \right. \\ & \left. + \Delta_2 \nu_{ij}^{(2)} \left( f_{i\uparrow}^{\dagger} f_{j\downarrow}^{\dagger} - f_{i\downarrow}^{\dagger} f_{j\uparrow}^{\dagger} - f_{i\uparrow} f_{j\downarrow} + f_{i\downarrow} f_{j\uparrow} \right) \right\} . \end{aligned} \tag{14}$$

The initial $N\lambda_3$ acts as an overall chemical potential and can be dropped. The following basis is then block-diagonal:

$$\hat{H}_{\text{QSL}} = \sum_{ij} \left( \begin{pmatrix} f_{i\uparrow}^{\dagger} & f_{i\downarrow} \end{pmatrix} \quad \begin{pmatrix} f_{i\uparrow} & f_{i\downarrow}^{\dagger} \end{pmatrix} \right) \begin{pmatrix} (h_{ij}) & (0) \\ (0) & (-h_{ij}) \end{pmatrix} \begin{pmatrix} \begin{pmatrix} f_{j\uparrow} \\ f_{j\downarrow}^{\dagger} \end{pmatrix} \\ \begin{pmatrix} f_{j\uparrow}^{\dagger} \\ f_{j\downarrow} \end{pmatrix} \end{pmatrix} , \tag{15}$$

where

$$h_{ij} = \begin{pmatrix} \frac{\lambda_3}{2}\delta_{ij} + t_a \nu_{ij}^{a} & \frac{\lambda_1}{2}\delta_{ij} + \Delta_2 \nu_{ij}^{(2)} \\ \frac{\lambda_1}{2}\delta_{ij} + \Delta_2 \nu_{ij}^{(2)} & -\frac{\lambda_3}{2}\delta_{ij} - t_a \nu_{ij}^{a} \end{pmatrix} \tag{16}$$

(a sum over $a = 1, 2$ is implicit). Hence, all the information is contained in the upper matrix:

$$\hat{H}_{\text{QSL}}^{U} = \sum_{ij} \begin{pmatrix} f_{i\uparrow}^{\dagger} & f_{i\downarrow} \end{pmatrix}_{\alpha} h_{ij}^{\alpha\beta} \begin{pmatrix} f_{j\uparrow} \\ f_{j\downarrow}^{\dagger} \end{pmatrix}_{\beta} . \tag{17}$$

The self-consistency conditions are:

$$\Delta_{ij} = -2\langle f_{i\uparrow} f_{j\downarrow} \rangle = 2\langle f_{i\downarrow} f_{j\uparrow} \rangle, \tag{18}$$

$$t_{ij} = 2\langle f_{i\uparrow}^{\dagger} f_{j\uparrow} \rangle = 2\langle f_{i\downarrow}^{\dagger} f_{j\downarrow} \rangle, \tag{19}$$

$$0 = \langle f_{i\uparrow} f_{j\uparrow} \rangle = \langle f_{i\downarrow} f_{j\downarrow} \rangle = \langle f_{i\uparrow}^{\dagger} f_{j\downarrow} \rangle = \langle f_{i\downarrow}^{\dagger} f_{j\uparrow} \rangle. \tag{20}$$

The physical Hilbert space, global half-filling, is enforced by the Lagrange multipliers $\lambda_1$ and $\lambda_3$:

$$\lambda_1 : 0 = \sum_{i} \langle f_{i\uparrow} f_{i\downarrow} \rangle - \langle f_{i\downarrow} f_{i\uparrow} \rangle, \tag{21}$$

$$\lambda_3 : 1 = \sum_{i} \langle f_{i\uparrow}^{\dagger} f_{i\uparrow} \rangle + \langle f_{i\downarrow}^{\dagger} f_{i\downarrow} \rangle. \tag{22}$$

This must now be diagonalized:

$$\hat{H}_{\text{QSL}}^{U} = \sum_{ij} \begin{pmatrix} \gamma_{i1}^{\dagger} & \gamma_{i2}^{\dagger} \end{pmatrix} D_{ij} \begin{pmatrix} \gamma_{j1} \\ \gamma_{j2} \end{pmatrix} \tag{23}$$

with diagonal $D$, and

$$h = UDU^{\dagger} , \tag{24}$$

$$\begin{pmatrix} f_{i\uparrow} \\ f_{i\downarrow}^{\dagger} \end{pmatrix} = U_{ij} \begin{pmatrix} \gamma_{j1} \\ \gamma_{j2} \end{pmatrix} = \begin{pmatrix} U_{ij}^{11}\gamma_{j1} + U_{ij}^{12}\gamma_{j2} \\ U_{ij}^{21}\gamma_{j1} + U_{ij}^{22}\gamma_{j2} \end{pmatrix} , \tag{25}$$

and the Hermitian conjugate gives the other required terms:

$$\begin{pmatrix} f_{i\uparrow}^{\dagger} & f_{i\downarrow} \end{pmatrix} = \begin{pmatrix} U_{ij}^{11*}\gamma_{j1}^{\dagger} + U_{ij}^{12*}\gamma_{j2}^{\dagger} & U_{ij}^{21*}\gamma_{j1}^{\dagger} + U_{ij}^{22*}\gamma_{j2}^{\dagger} \end{pmatrix}. \tag{26}$$

In this basis,

$$\{\gamma_{i\alpha}, \gamma_{j\beta}\} = \{\gamma_{i\alpha}^{\dagger}, \gamma_{j\beta}^{\dagger}\} = 0, \tag{27}$$

$$\{\gamma_{i\alpha}, \gamma_{j\beta}^{\dagger}\} = \delta_{ij}\delta_{\alpha\beta} , \tag{28}$$

$$\langle \gamma_{i\alpha}^{\dagger} \gamma_{j\beta} \rangle = \delta_{ij}\delta_{\alpha\beta} n_{\text{D}}(D_{ii}^{\alpha\alpha}), \tag{29}$$

where $n_\mathrm{D}$ is the Fermi-Dirac distribution. However, note that the eigenvalues $D_{ii}$ are ordered low to high, and the spectrum is symmetric about zero. Hence, at $T \approx 0$ (since $T^*/J_\mathrm{K} = 0.26\ \mathrm{K}/190\ \mathrm{K} \ll 1$), $n_\mathrm{D}(D_{mm}^{22}) = 0$, $n_\mathrm{D}(D_{mm}^{11}) = 1$. Feeding these expressions into the self-consistency conditions gives

$$\Delta_2 v_{ij}^{(2)} = 2\sum_m U_{im}^{11}\,U_{jm}^{21*}\,, \tag{30}$$

$$t_a v_{ij}^{(a)} = 2\sum_m U_{im}^{11*}\,U_{jm}^{11}\,, \tag{31}$$

$$\lambda_1 : 0 = \sum_{im} U_{im}^{12}\,U_{im}^{22*} - U_{im}^{21*}U_{im}^{11}\,, \tag{32}$$

$$\lambda_3 : 1 = \sum_i \left|U_{ii}^{11}\right|^2 + \left|U_{ii}^{22}\right|^2\,. \tag{33}$$

We set $t_1 = 1$, defining the energy scale. Working in $q$-space at $q = 0$ (since the gap should be constant), we found a self-consistent solution with

$$\Delta_2 = 0.4583, \tag{34}$$

$$t_2 = -0.2849, \tag{35}$$

$$\lambda_1 = 0.4327, \tag{36}$$

$$\lambda_3 = 1.500. \tag{37}$$

We used a tolerance of $10^{-3}$ in finding the constraints with the Lagrange multipliers:

$$\lambda_1 \Rightarrow \sum_{im} U_{im}^{12}\,U_{im}^{22*} - U_{im}^{21*}U_{im}^{11} = -8.4\times10^{-4}\ (\approx 0), \tag{38}$$

$$\lambda_3 \Rightarrow \sum_i \left|U_{ii}^{11}\right|^2 + \left|U_{ii}^{22}\right|^2 = 1.001\ (\approx 1). \tag{39}$$

The overall energy gap (identified from the density of states)

$$2\Delta = 0.44t_1 = 33\ \mathrm{K} \tag{40}$$

is essentially equal to the gap ($0.43t_1$) identified previously using exact diagonalisation[40].

In the witness-witness spin interactions $J_{ij}$ in Eqns. (1-4), the only free parameter is $\gamma$. We constrain $|\gamma| = 60\ \mathrm{K} \approx J_\mathrm{K}/3$ by requiring a match to the widely reported experimental value of the Curie-Weiss temperature $\theta_\mathrm{CW}(1\ \mathrm{K} < T) = -1.1\ \mathrm{K}$.

One plausibly estimates[42,43]

$$r_0 = \frac{\hbar v_\mathrm{F}}{2\Delta}, \tag{41}$$

where the spinon band structure enters via the gap $2\Delta$, and the spinon Fermi velocity $v_{\mathrm{F}}$ of the parent U(1)$[0,\pi]$ gapless QSL from which the $\mathrm{Z}_2[0,\pi]\beta$ forms[29] with

$$v_{\mathrm{F}} = \frac{\sqrt{2}t_1 d}{\hbar}\,, \tag{42}$$

where $d$ is the nearest-neighbor kagome Cu spacing. Eqns. (41,42) lead to Eqn. (6).

## D. Witness spin Monte Carlo simulations

In herbertsmithite, the witness spin sites (i.e. $Zn^{2+}$ sites) form a triangular lattice on the $ab$-plane, staggered along the $c$-axis with a period of 3. This witness spin lattice effectively connects as a simple cubic lattice[17]. We simulate a witness spin lattice with the size of $X \times X \times Z = 45 \times 45 \times 4$. The smallest cell containing one witness spin, which is $1 \times 1 \times 1$ under this notation, is a rhombic prism with the side length $\frac{a}{\sqrt{3}} = 3.95$ Å and height $c = 14.09$ Å. The direction of its rhombic base is rotated by 90° around the $c$-axis, compared to the rhombic base of the conventional unit cell of herbertsmithite. To satisfy periodic boundary conditions, $X$ has to be a multiple of three and $Z$ has to be an even number.

We created a Monte Carlo simulation of the witness spins using the Metropolis-Hasting algorithm. We modelled the witness spins as classical Ising spins $s_i = \pm 1/2$. Although witness spins in herbertsmithite are not Ising-like, they are not Heisenberg-like, either. Electron spin resonance demonstrates a strong Dzyaloshinskii–Moriya (DM) interaction ($D/J = 0.08$) leading to spin anisotropy[46]. Moreover, there is evidence for additional easy-axis anisotropy beyond the DM interaction[66]. A key consequence of this magnetic anisotropy is that using a pure Heisenberg representation to model witness spin dynamics would be incorrect and that using an Ising representation is a reasonable approximation choice. The use of Ising spins has a further pragmatic justification in the context of spin glass. Adding a tiny amount of anisotropy to a Heisenberg spin glass can lead to a spin glass in the Ising universality class, making Ising-spin model effective in reproducing experimental observations[67]. We initialized the system with the size 45×45×4 and periodic boundary conditions with 33% of potential witness spin sites occupied ($N = 2763$ spins). The configuration of occupied sites is randomly assigned using a seeded random number generator. We average all of our results over 128 different configurations, keeping the same sets of seed across all runs. For each seeded configuration, we also average our results over 3 simulation runs.

In all cases, the initial state of the system has each spin in random uncorrelated states, corresponding to infinite temperature. We then run our simulation starting at $T = 400$ mK — above the freezing transition temperature so as to avoid quenching the glass — and ending at 50 mK at intervals of 10 mK. In addition, we also conduct a separate run going from 10 K to 1.6 K at intervals of 0.2 K to ensure that the Curie-Weiss temperature is $\theta_{\mathrm{CW}} = -1.1$ K, and from 1.5 K to 0.5 K at intervals of 0.1 K for completeness.

At each temperature point, we first equilibrate the system by updating the system over 1000 sweeps, with each sweep consisting of $N = 2763$ update steps. We then sampled the spin-per-site noise $s(t) = \frac{1}{N}\sum_i s_i(t)$, Edwards-Anderson spin-glass order parameter $q_{\mathrm{EA}} = \frac{1}{N}\sum_i \overline{2s_i(t)}^2$, and antiferromagnetic order parameter $\phi_{\mathrm{AF}} = \overline{\frac{1}{N}\left(\sum_i (-1)^k (2s_i(t))\right)^2}$ ($k = 0,1$ for each sublattice of the bipartite witness spin sites) over 100000 sweeps. The bar represents an average over the Monte Carlo sweep time. From $s(t)$, the magnetization noise $M(t)$ and DC magnetic susceptibility $\chi$ are estimated using

$$M(t) = \rho_V \mu_0 g \mu_{\mathrm{B}} s(t) \sqrt{\frac{N}{N_{\mathrm{EXP}}}}, \tag{43}$$

$$\chi = \rho_V \mu_0 (g\mu_{\mathrm{B}})^2 N \frac{\overline{s^2(t)} - \left(\overline{s(t)}\right)^2}{k_{\mathrm{B}} T}, \tag{44}$$

where $\rho_V$, $\mu_0$, $g = 2$, $\mu_{\mathrm{B}}$, $k_{\mathrm{B}}$ are the number density per volume of witness spins in herbertsmithite (33% per Zn sites), vacuum permeability, electron $g$-factor, Bohr magneton, and Boltzmann constant, respectively. The factor $\sqrt{N/N_{\mathrm{EXP}}}$, where $N_{\mathrm{EXP}}$ is the number of witness spins in the volume of herbertsmithite Sample 1 (~3 $\mathrm{mm}^3$), is required to approximately estimate the order of the magnetization noise magnitude that generally scales as $M(t) \propto 1/\sqrt{N_{\mathrm{EXP}}}$ [33]. Error bars of $\chi$, $\phi_{\mathrm{AF}}$, $q_{\mathrm{EA}}$ are the standard error of averaging.

The predicted witness spin magnetization noise $M(t)$ is then processed in the same method as the experimental spin noise in $ZnCu_3(OH)_6Cl_2$ (Methods B). The distribution is Gaussian with small statistical fluctuations (Extended Data Fig. 4a). The PSD in Figs. 4c,d and Extended Data Fig. 4b has a frequency resolution $\Delta\omega = (2\pi\ \mathrm{rad}) \times (10^{-5}/\mathrm{MCS}) = 6 \times 10^{-5}$ rad/MCS and is averaged over a $10\Delta\omega$ or $100\Delta\omega$ window at high frequencies; error bars are the standard error of averaging. The power-index in Fig. 5a is obtained by fitting $6 \times 10^{-5}$ rad/MCS $\leq \omega \leq 1 \times 10^{-3}$ rad/MCS, as shown in Extended Data Fig. 4b; error bars are the standard error from fitting. In Fig. 5a, open symbol points are used at temperatures above $T^*$ where the power-law fitting is challenging ($R$-value < 0.98). The variance in Fig. 5b is calculated by integrating the PSD from $6 \times 10^{-5}$ rad/MCS $\leq \omega \leq 6 \times 10^{-2}$ rad/MCS. The variance peak temperature remains at $T^*$ for different integration ranges (Extended Data Fig. 4c). When 1 Monte-Carlo step is set to 1 MCS = 100 μs, the simulated PSD and the measured experimental PSD roughly correspond in the same frequency window (Extended Data Fig. 5). They remain consistent with each other for any values in the range 1 MCS < 100 μs, as long as the PSD continues to be scale-invariant down to lower frequency both in simulation and experiment. The challenges in fitting power spectral density at temperatures above $T^*$, which were not seen in experiment, may be resolved by simulating power spectral density in a lower Monte-Carlo-frequency range or by simulating the fully quantum mechanical theory.

When we perform the equivalent simulations for different witness spin concentrations from 15% to 60%, the transition temperature $T^*$ changes from 200 mK to 100 mK, and the nearest-neighbor witness spin interaction energy scale (Eqn. (7)) changes from 2.5 K to 0.5 K. Even if these different witness spin concentrations are used in the model, these quantitative changes do not alter the conclusion of this work.

We finally note that whether the predicted transition is of true spin-glass type has not been examined in detail. Direct theoretical investigations into this point, such as finite-size scaling of the spin-glass susceptibility, are left for future work.

### E. Neutron scattering structure factor

Extended Data Fig. 6 shows the witness spin structure factor $\Sigma(\boldsymbol{q})$ that is calculated from a spin-configuration snapshot in our Monte Carlo simulation at 2 K over 1000 sweeps and averaging over 128 configurations.

$$\Sigma(\boldsymbol{q}) = |F(\boldsymbol{q})|^2 \left\langle \left| \sum_i s_i e^{i\boldsymbol{q}\cdot\boldsymbol{r}_i} \right|^2 \right\rangle, \tag{45}$$

where $F(\boldsymbol{q})$ is the magnetic form factor of $Cu^{2+}$. It reasonably matches the low-energy neutron scattering structure factor in $ZnCu_3(OH)_6Cl_2$, which shows diffuse scattering without a sharp peak[17,27] albeit with a quite low signal-to-noise ratio. Comparable features are the shape of the mid-intensity contribution (green) that extends throughout the in-plane and out-of-plane directions, the high-intensity contribution (red) at the out-of-plane peak at $(00\frac{3}{2})$, and the high-intensity in-plane circular shape contribution with correct $|\boldsymbol{q}|$ and approximately equally distributed intensity. Further comparison of the precise shape of the experimental neutron scattering intensity pattern in the $(HK0)$ plane requires improvement both in the precision of the neutron scattering experiment and in the model.

### F. Considering alternatives to spinon-mediated witness spin interactions

While we considered a large number of alternative hypotheses, the only cases capable of explaining the full range of experimental data involved spinon-mediated couplings via spin liquids. In reviewing these hypotheses there are two strong constraints:

First, any theoretical model with sufficiently *rapid* variation with distance of the witness spin to witness spin interaction decay (local couplings) will result in a significant population of isolated witness spins. With Zn site-occupation probability $p$, the percentage of isolated witness spins having no nearest neighbour is $(1-p)^6$; with $p = 0.33$ this gives 9% (3% of Zn sites). These isolated witness spins must contribute a DC magnetic susceptibility which diverges as $1/T$ as $T \to 0$. As shown in Extended Data Fig. 7a, if only 0.7% of Zn sites are occupied by isolated witness spins, this would be enough to show a DC susceptibility evolving as $1/T$ as $T \to 0$ which is not observed in any of our experiments.

Second, witness spin to witness spin interactions evolving too *slowly* with distance ($1/r^2$ or slower) will lead to an unphysical divergence in the sum over spins forming the structure factor. As well as being unphysical, this situation is incompatible with the experimental inelastic neutron scattering (INS) structure factor which shows broad features in momentum space at 2 K consistent with dominant antiferromagnetic nearest-neighbour correlations[17]. An example of the structure factor for a slowly decaying spin-wave mediated interaction is shown in Extended Data Figs. 7b,c.

In the context of these constraints, we have considered and ruled out a variety of alternatives to spinon-mediated witness spin interactions including nearest neighbor local exchange, next-nearest neighbor local exchange, direct dipolar witness spin interactions, dimer correlations mediating the witness spin interaction, random singlets and random spin clusters, and spin-wave mediating the witness spin interaction, as discussed in the Supplementary Discussion.

The model in the main text discusses $Z_2[0,\pi]\beta$ QSL scenario, while another candidate kagome QSL in herbertsmithite is the $U(1)[0,\pi]$ state with a Dirac nodal spinon Fermi surface. We calculated its spinon band structure using the mean field decoupling of ref. 29, then the witness spin interaction using the same methods we used for $Z_2$: in this case, the calculation becomes that of an RKKY interaction between witness spins mediated by the spinon Dirac-nodal Fermi surface. We find that all witness couplings are AF, decaying approximately as $1/r^3$. The predictions of the witness spin noise, susceptibility, and order parameters via these U(1) QSL spinon-mediated witness spin interactions are shown in Extended Data Fig. 8. Each panel can be compared to Figs. 4,5. The U(1) model predicts a transition at 110 mK, which is smaller than the 150 mK of the $Z_2$ quantum spin liquid prediction. Thus, our $Z_2$ quantum spin liquid model is more consistent with the experiment. However, the U(1) quantum spin liquid model also reproduces the qualitative features of the experiment in Figs. 2,3, and cannot be fully excluded using the existing data.

**Data Availability:** Data presented in this paper are deposited in Zenodo https://doi.org/10.5281/zenodo.15114443 (ref. 68). Source data are provided with this paper.

**Code Availability:** Simulation codes used in this paper are deposited in Zenodo https://doi.org/10.5281/zenodo.15114443 (ref. 68).

Correspondence and requests for materials should be addressed to Felix Flicker or J. C. Séamus Davis.

**Methods-only references**

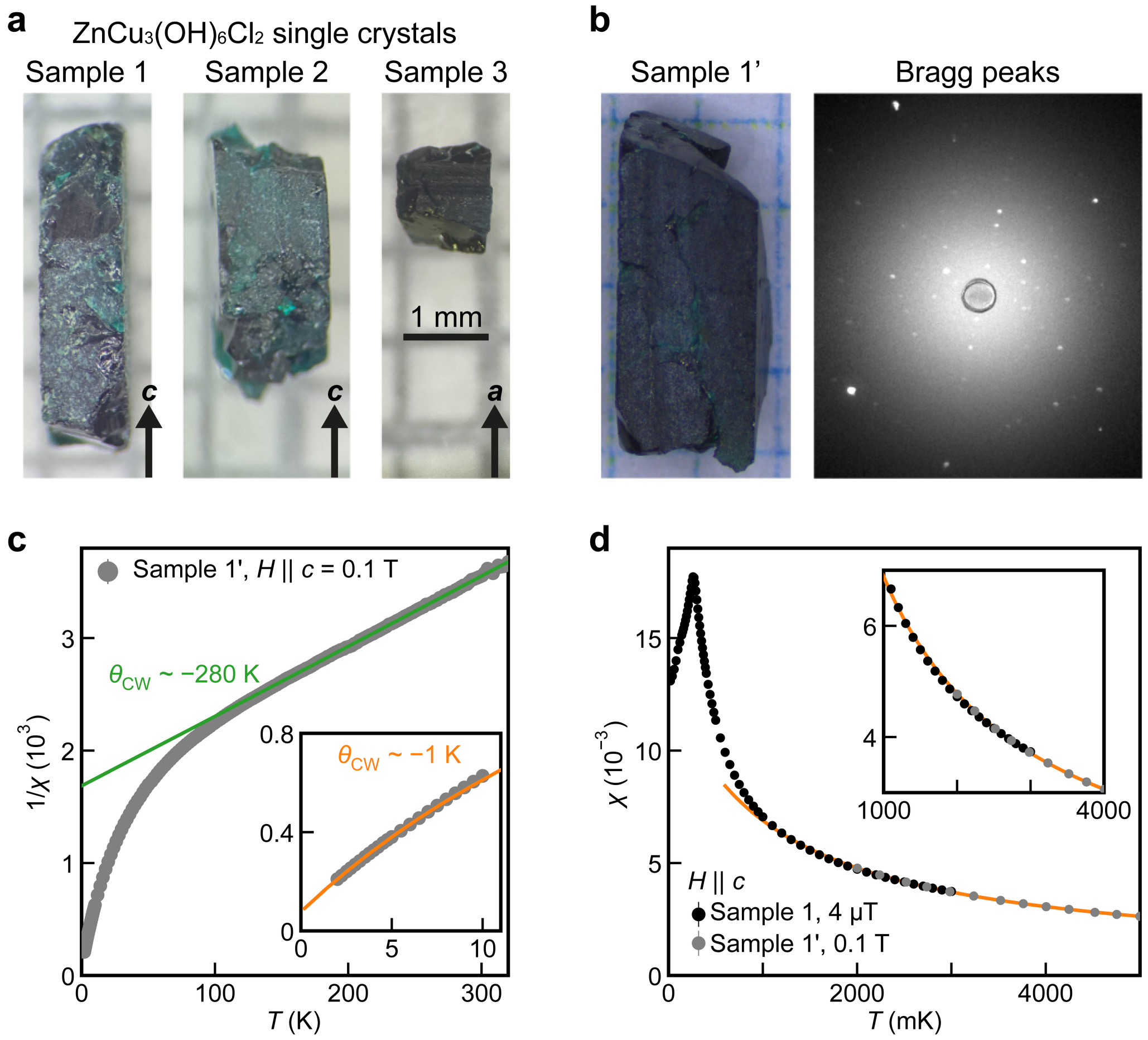

**Extended Data Fig. 1 Single crystals of $ZnCu_3(OH)_6Cl_2$ and their characterization**

**a.** Photos of three $ZnCu_3(OH)_6Cl_2$ single crystals. Sample 1 is obtained by polishing down Sample 1' in **b**. Measurements were performed along the *c*-axis for Samples 1 and 2, and along the *a*-axis for Sample 3, as indicated by the arrows.

**b.** Clear Bragg peaks of the x-ray Laue diffraction measured for Sample 1'.

**c.** DC susceptibility of Sample 1' in SI units measured by a SQUID MPMS (Quantum Design). The solid lines represent Curie-Weiss fittings by $\chi = \chi_0 + \frac{C_{\mathrm{Curie}}}{T-\theta_{\mathrm{CW}}}$. Fitting in a temperature range 150 K $\leq T \leq$ 320 K (green) yields $\chi_0 = -5 \times 10^{-6}$, $\theta_{\mathrm{CW}} = -280$ K, and $C_{\mathrm{Curie}} = 0.165$ K and that in a temperature range 2 K $\leq T \leq$ 6 K (orange, in inset) yields $\chi_0 = 4.3 \times 10^{-4}$, $\theta_{\mathrm{CW}} = -1.07$ K, and $C_{\mathrm{Curie}} = 0.0134$ K.

**d.** The overlay of the micro-Tesla susceptibility in Fig. 3c (black) and sub-Tesla susceptibility in **c** (gray). The inset magnifies the overlapping temperature range. The fitting in the range 2 K $\leq T \leq$ 6 K (orange) captures the susceptibility down to 1 K.

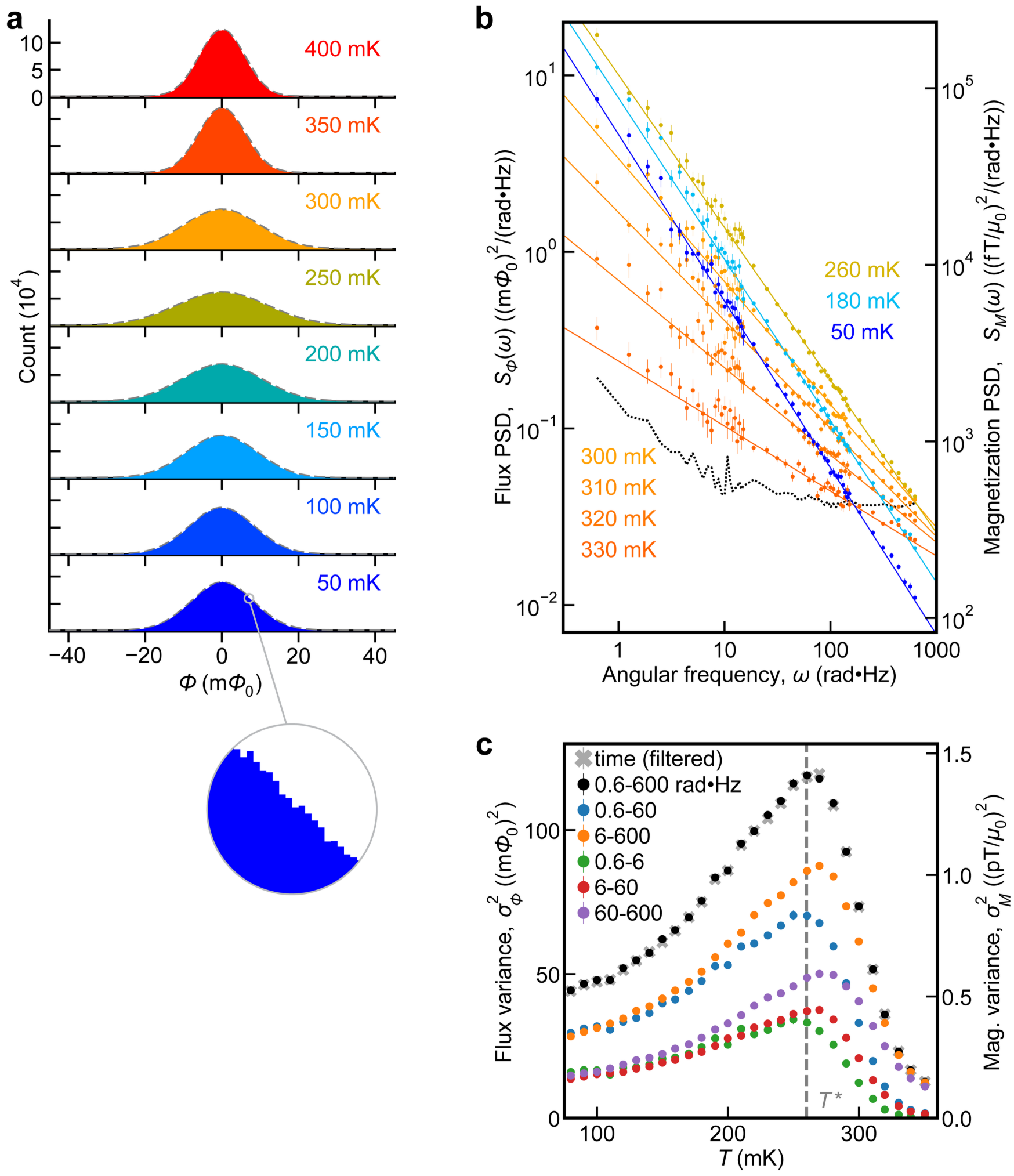

**Extended Data Fig. 2 Analysis of $ZnCu_3(OH)_6Cl_2$ spin noise spectroscopy**

**a.** Typical distribution histograms of the measured $ZnCu_3(OH)_6Cl_2$ witness spin flux noise $\Phi(t,T)$ at eight selected temperatures (frequency components out of bandwidth 0.3 rad•Hz $\leq \omega \leq$ 600 rad•Hz are filtered out). Each histogram follows a Gaussian distribution (gray dashed line), with small statistical fluctuations (inset).

**b.** Measured witness spin flux noise power spectral density $S_\Phi(\omega,T)$. Error bars are the standard error of segment averaging. The black dotted line indicates the measured background contribution that is subtracted from the signal. Lines represent $A(T)\omega^{-\alpha(T)}$ fitting in the frequency range 0.6 rad•Hz $\leq \omega \leq$ 600 rad•Hz.

**c.** Comparison of measured witness spin flux noise variance $\sigma_\Phi^2$ obtained by direct calculation from $\Phi(t,T)$ (after filtering) and by integrating $S_\Phi(\omega,T)$ in different frequency ranges. Error bars are propagated from power spectral density.

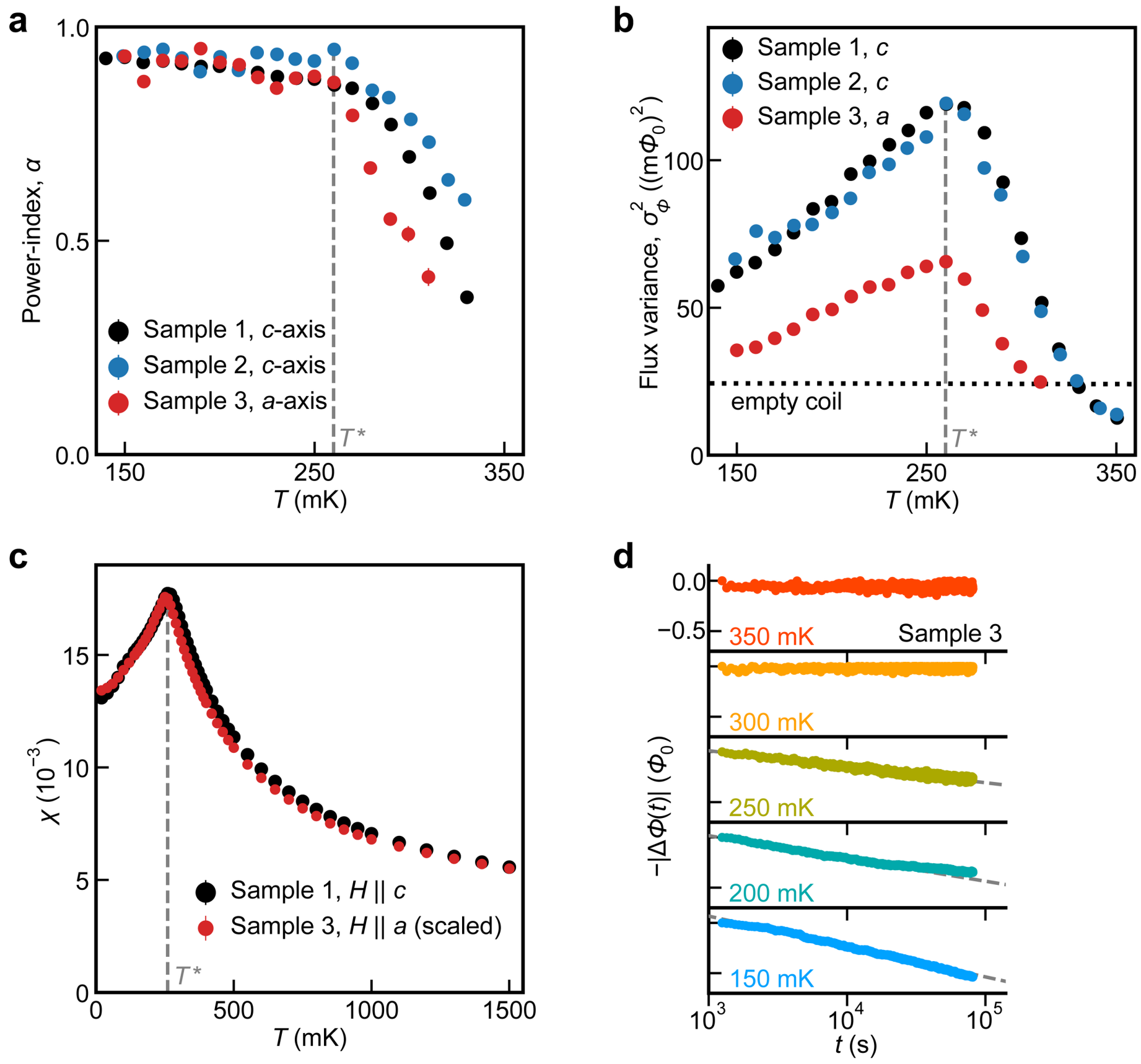

**Extended Data Fig. 3 Comparison of spin noise and susceptibility measurements in multiple $ZnCu_3(OH)_6Cl_2$ single crystals**

**a.** Measured witness spin noise power-index $\alpha$ from $S_\Phi(\omega, T) \propto \omega^{-\alpha(T)}$ as a function of temperature. Error bars are the standard error from fitting. A clear transition to $\alpha \approx 1$ is detected at $T^* \approx 260$ mK in all samples.

**b.** Measured witness spin noise variance $\sigma_\Phi^2$ as a function of temperature. A transition in noise power at $T^* \approx 260$ mK (dashed line) is observed in all samples. The dotted line indicates the background noise variance.

**c.** Measured witness spin susceptibility $\chi(T)$ (SI units) in micro-Tesla magnetic fields, showing a sharp transition to a rapidly diminishing $\chi(T)$ below $T^* \approx 260$ mK. The result of Sample 3 is scaled to facilitate a comparison.

**d.** Time evolution of Sample 3 average flux $\Phi(t)$ against a 2 $\mu$T applied field after a temperature quench from a thermalized condition at 400 mK, showing a $-\ln t$ relaxation (dashed line) on periods of a day only below $T^* \approx 260$ mK. The same behavior of Sample 1 is shown in the Fig. 3d inset.

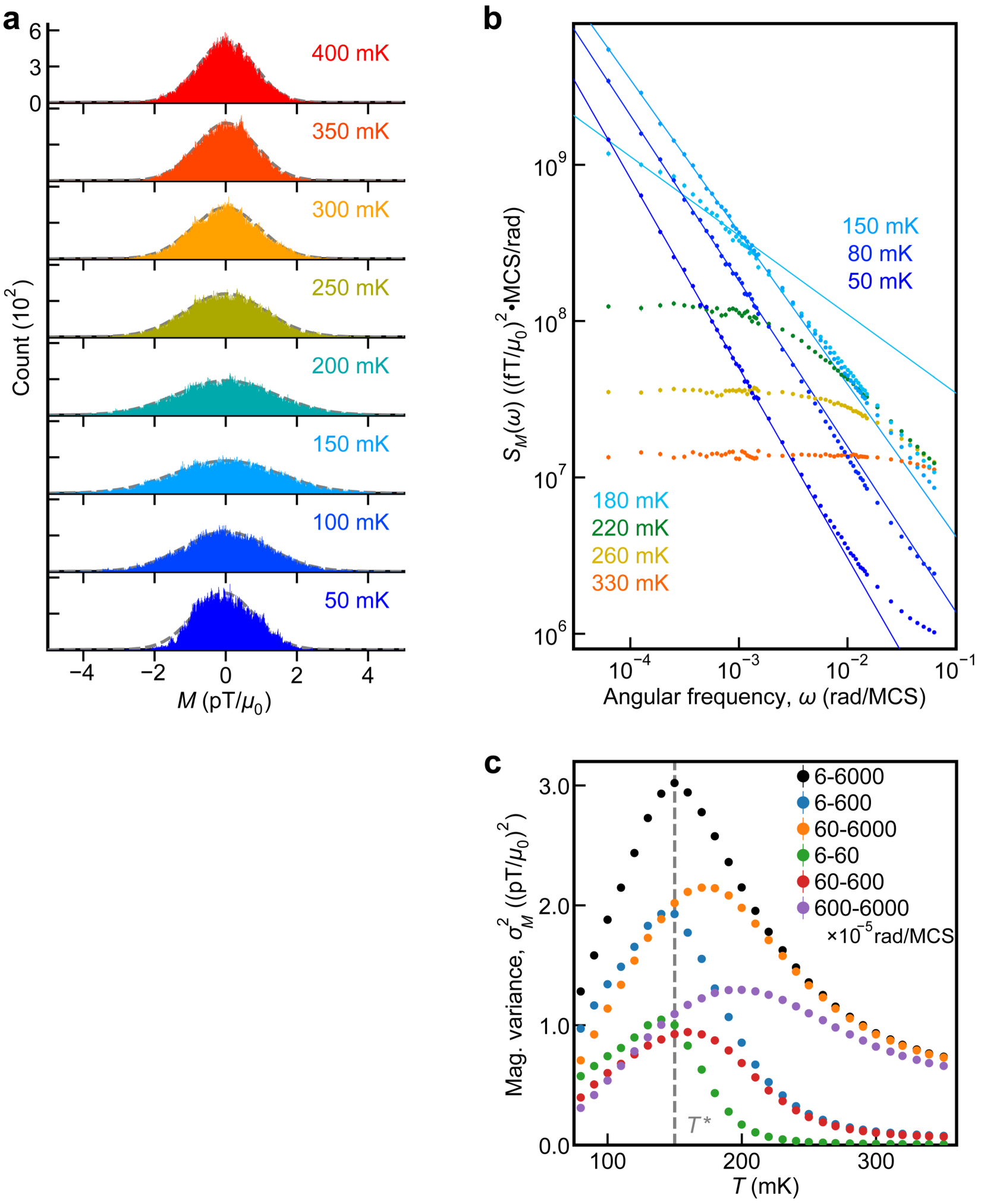

**Extended Data Fig. 4 Analysis of simulated spinon-mediated witness spin noise spectroscopy**

**a.** Typical distribution histograms of the simulated witness spin magnetization noise $M(t,T)$ at eight selected temperatures (frequency components out of bandwidth $3 \times 10^{-5}$ rad/MCS $\leq \omega \leq 6 \times 10^{-2}$ rad/MCS are filtered out). Each histogram follows a Gaussian distribution (gray dashed line) with small statistical fluctuations.

**b.** Predicted witness spin magnetization noise power spectral density $S_M(\omega,T)$. Error bars are the standard error of Monte-Carlo-run averaging. Lines represent $A(T)\omega^{-\alpha(T)}$ fitting in the frequency range $6 \times 10^{-5}$ rad/MCS $\leq \omega \leq 1 \times 10^{-3}$ rad/MCS.

**c.** Comparison of simulated witness spin magnetization noise variance $\sigma_M^2$ obtained by integrating $S_M(\omega,T)$ in different frequency ranges.

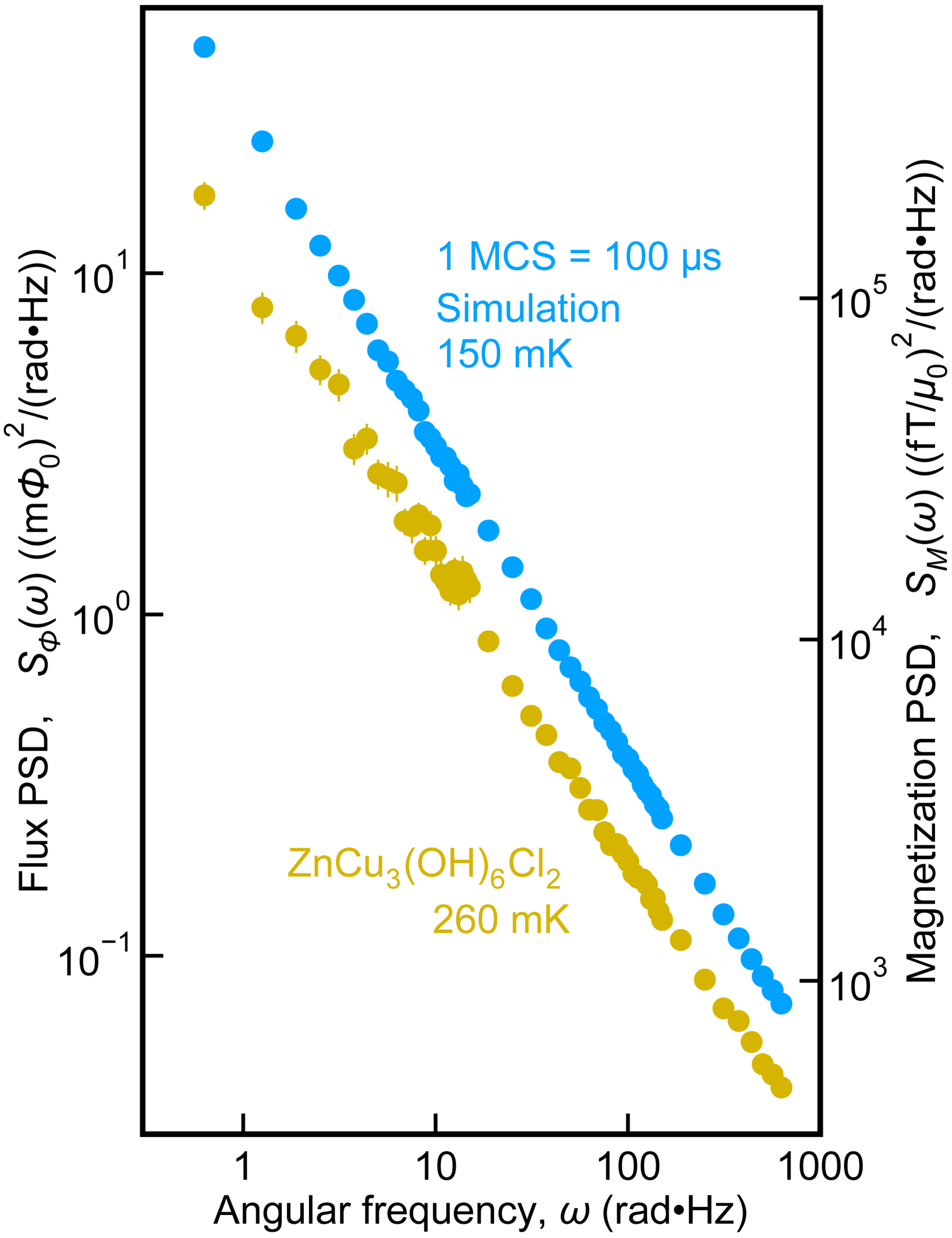

**Extended Data Fig. 5 Witness spin noise power spectral density when 1 Monte Carlo step = 100 μs**

Simulated witness spin magnetization noise power spectral density $S_M(\omega)$ at $T^*$ (blue), when 1 Monte Carlo step is set to 100 μs. It falls into the frequency window of measured $ZnCu_3(OH)_6Cl_2$ witness spin noise at $T^*$ (yellow): 0.6 rad•Hz $\leq \omega \leq$ 600 rad•Hz. Error bars are the standard error of segment averaging.

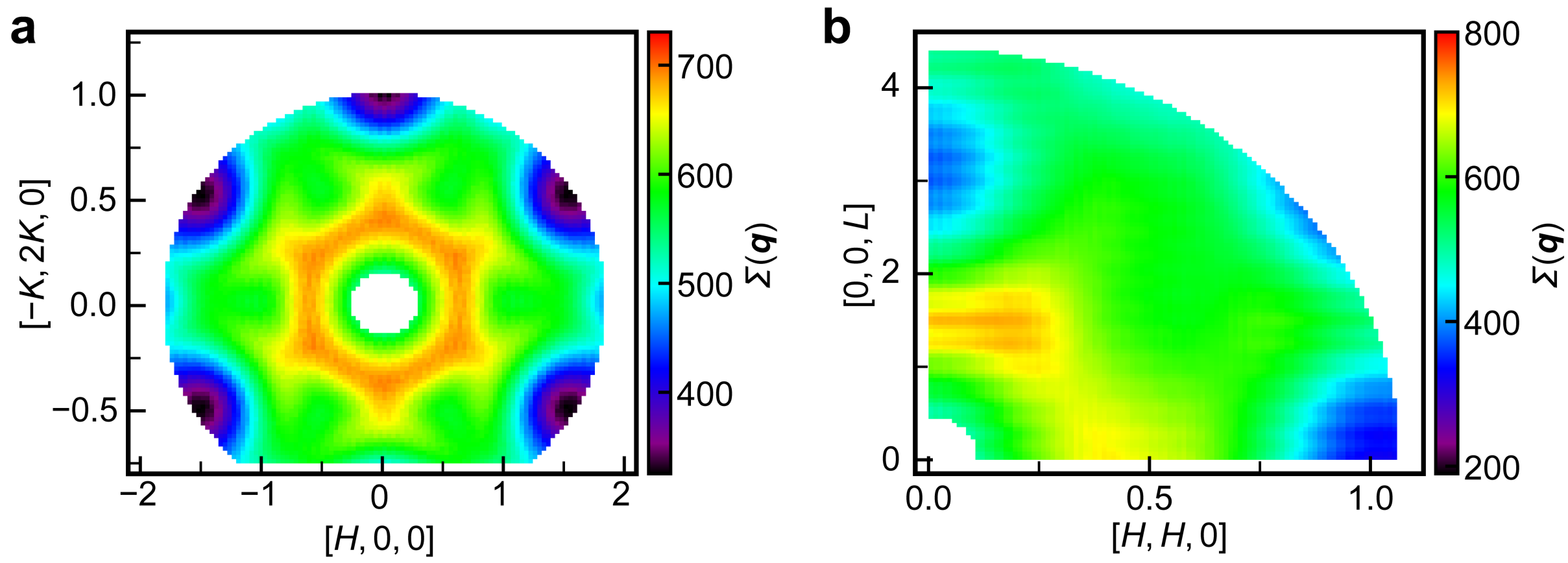

**Extended Data Fig. 6 Simulated witness spin structure factor**

The witness spin structure factor $\Sigma(\boldsymbol{q})$ that is calculated from a spin-configuration snapshot in the Monte Carlo simulation at 2 K for (*HK0*) space (**a**) and (*HHL*) space (**b**).

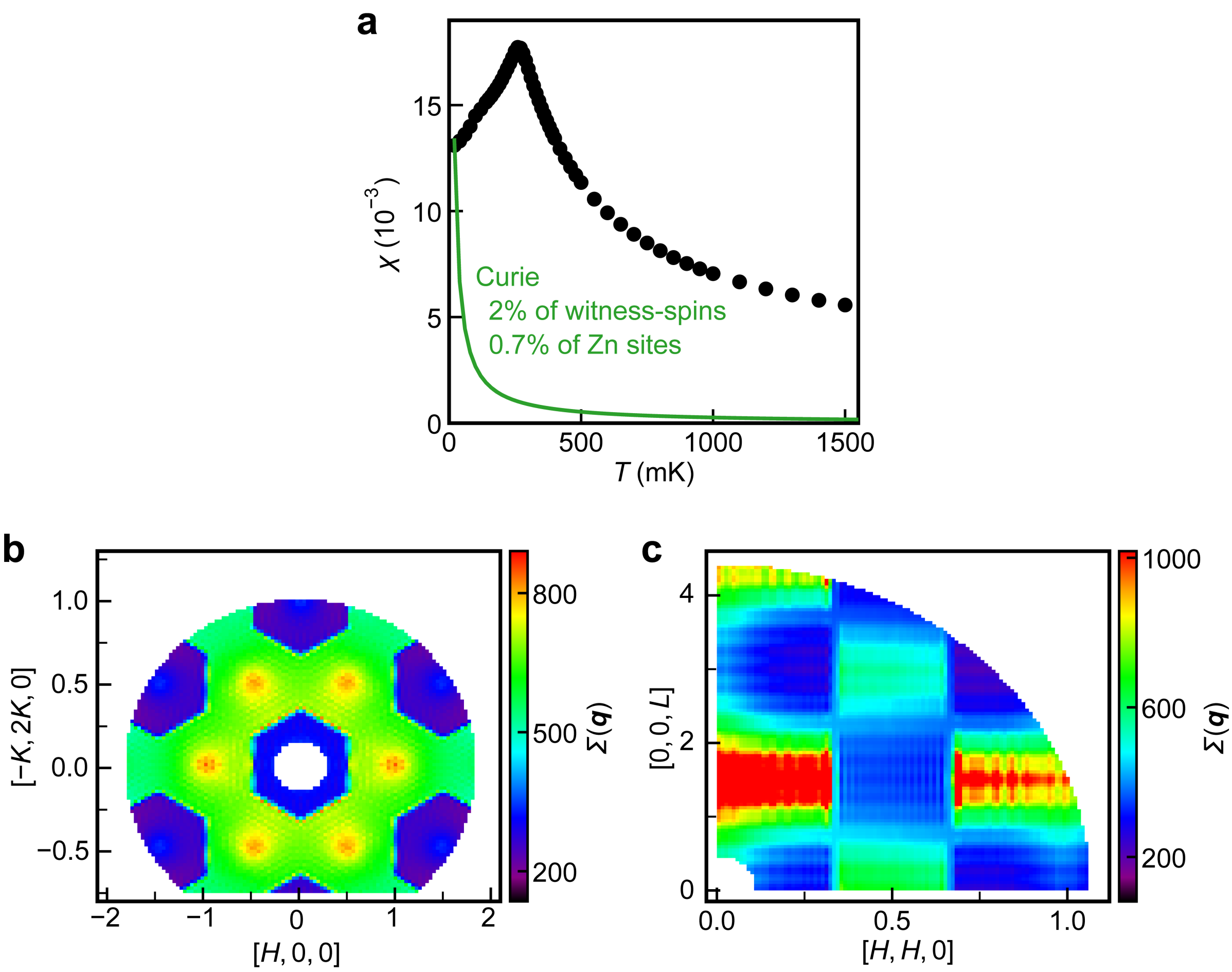

**Extended Data Fig. 7 Observations constraining alternative theoretical models**

**a.** The estimated isolated-spin magnetic susceptibility $\chi(T)$ of 2% of witness spins (0.7% of Zn sites) overlayed on the measured witness spin susceptibility of Fig. 3c (black dots).

**b,c**. The witness spin structure factor $\Sigma(\boldsymbol{q})$ assuming spin-wave mediated interactions that decay as $1/r \sim 1/r^2$. It is calculated from a snapshot of witness spin configuration in the Monte Carlo simulation at 2 K for (HK0) space (**b**) and (HHL) space (**c**). The quantitative detail is dependent on the in-plane distance cutoff due to the slow decay of the interaction, and here a 45x45 cutoff is used.

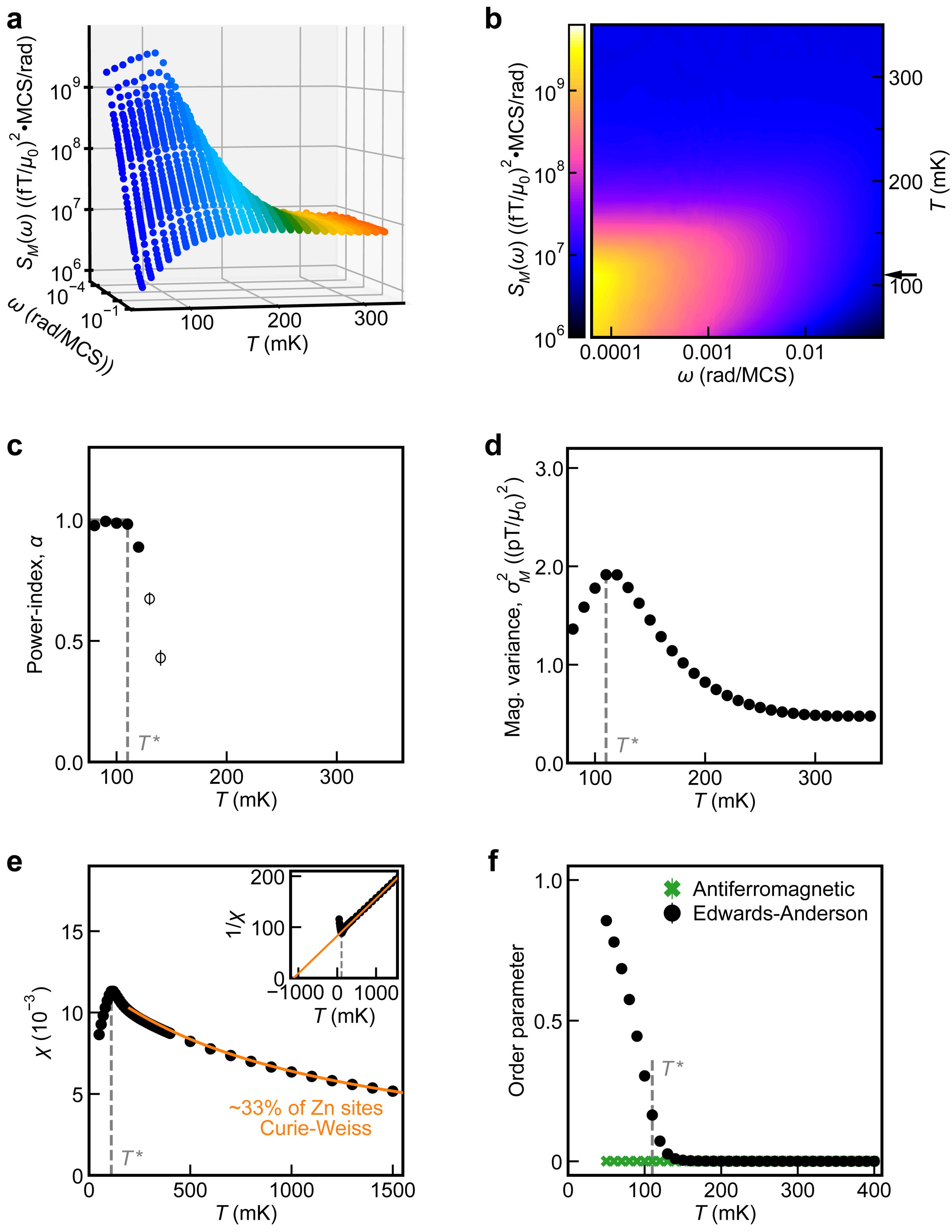

**Extended Data Fig. 8 U(1) quantum spin liquid spinon-mediated witness spin dynamics in $ZnCu_3(OH)_6Cl_2$**

**a.** Predicted power spectral density of witness spin magnetization noise $S_M(\omega,T)$ as a function of frequency and temperature due to spinon-mediated interactions.

**b.** Contour plot of $S_M(\omega,T)$ from **a** revealing a clear transition in dynamics at $T^* \approx 110\ \mathrm{mK}$ (horizontal arrow).

**c.** Predicted witness spin magnetization noise power-index $\alpha$ for $S_M(\omega,T) \propto \omega^{-\alpha(T)}$ as a function of temperature from **a**, revealing a transition to $\alpha \approx 1$ at $T^*$ (dashed line). The open circles are used for temperatures above $T^*$ where a power-law fitting is challenging. Error bars are the standard error from fitting.

**d.** Predicted witness spin magnetization noise variance $\sigma_M^2$ as a function of temperature from **a**, indicating a transition in noise power at $T^*$ (dashed line).

**e.** Predicted witness-spin-only susceptibility $\chi(T)$ (SI units, inset shows $1/\chi$) due to spinon-mediated interactions, revealing a sharp transition at $T^*$ (dashed line) from a Curie-Weiss behaviour determined at higher temperatures (orange line).

**f.** Predicted antiferromagnetic order parameter (green crosses) and Edwards-Anderson spin-glass order parameter (black dots) from witness spin simulations, indicating that $T^*$ (dashed line) is a spin-freezing transition.

## Supplementary Discussion:

### A. Hypothetical alternatives to spinon-mediated witness spin interactions

1. *Nearest neighbor (NN) local exchange*

Antiferromagnetic (AF) NN witness spin interactions could occur via some sequence of exchange pathways[17]. The shortest path would be Cu-O-O-Cu (super-super exchange) connecting two witness planes. However, this NN case is un-frustrated for either ferromagnetic (FM) or AF interactions, and so cannot account for the observed spin glass transition at $T^*$. Models of random NN exchange with a Gaussian distribution of values centered on the net AF interaction strength (1 K) seen in neutron scattering, give rise to long-range AF order; we confirmed this by simulation. Finally, any NN model would also leave $\sim$9% of isolated witness spins in our samples, inconsistent with the observed DC magnetic susceptibility as $T \rightarrow 0$.

2. *Next-nearest neighbor (NNN) local exchange*

In principle, some as-yet unknown AF (frustrating) NNN exchange pathway might exist. However, there is no evidence of NNN correlations in neutron scattering[17]. Even if NNN interactions exist and are captured by an AF NN+NNN model, NN interactions would give FM order on triangular witness spin planes that are then AF coupled from plane to plane, and NNNs would then frustrate the in-plane FM order. The resulting order would likely be the standard 120-degree order on a witness spin plane which is AF coupled from plane to plane. To explore this concept we modeled a range of AF NNN interaction strengths from 0 to the NN strength in our MC simulations. We found that all simulations led to sizable AFM order parameters $\phi_{\mathrm{AF}} > 0.4$ (cf. $\phi_{\mathrm{AF}} < 0.02$ in Fig. 2d for the $Z_2$ predicted QSL model) suggesting significant long-range order, which is not detected in specific heat or neutron scattering measurements. Finally, any NN+NNN model would also leave $\sim$2.4% of isolated witness spins, again inconsistent with the observed DC magnetic susceptibility as $T \rightarrow 0$.

3. *Direct dipolar witness spin interactions*

While long-range dipolar witness spin interactions could be frustrating, the associated energy scale is too small. With a NN witness-witness distance of 6.1 Å, the NN dipole-dipole interaction energy between two $s = 1/2$ spins is

$$\frac{J_{\mathrm{NN}}}{k_{\mathrm{B}}} = \frac{1}{k_{\mathrm{B}}} \frac{\mu_0}{4\pi} \frac{(2\mu_{\mathrm{B}})^2}{\left(6.1\ \text{Å}\right)^3} = 11\ \mathrm{mK}, \tag{46}$$

which is incompatible with the observed $T^* = 260\ \mathrm{mK}$. Since NN interactions are unfrustrated, any predicted $T^*$ would rely on further neighbor couplings and so would be reduced even further below this estimated temperature.

4. *Dimer correlations mediating the witness spin interaction*

One can hypothesize effects of Zn substituted into the Cu Kagome plane[11,20,24,49,56]. Ref. 69 suggests that two spins on each triangle containing a Zn-substituted site may form an unfrustrated singlet (dimer). However, for such a hypothesis ref. 70 calculates the Knight shift around Zn-substitution and finds it to decay as $1/r$; such a slow decay is inconsistent with the experimental structure factor. Further, ref. 49 suggests that these disturbances lead to a Curie-like $1/T$ susceptibility, incompatible with the the observed DC magnetic susceptibility as $T \to 0$.

There is also a theoretical suggestion that a free-spin contribution might arise in the kagome layers due to Zn-substitution[71]. However, such a contribution is inconsistent with both the experimental structure factor which indicates that the witness spins are embedded in the Zn layers[17], and with the observed DC magnetic susceptibility as $T \to 0$.

5. *Random singlets and random spin clusters*

Power-law scaling in herbertsmithite has been identified in specific heat[53,72] and the dynamic spin susceptibility[26], which led to the hypothesis of a random singlet state arising from randomly distributed AF interactions. Ref. 73 studies random singlets and predicts, as a function of decreasing temperature, a magnetic susceptibility that increases as $1/T$, then plateaus, then increases as $1/T^a$ with $a < 1$. A recent study on an organic QSL candidate similarly found a low-$T$ upturn related to random singlets[74]. However, herbertsmithite shows no susceptibility divergence in the observed DC magnetic susceptibility as $T \rightarrow 0$.

Ref. 73 also discusses qualitatively that coexisting random FM & AF interactions could lead to a partial formation of FM spin clusters of random size. However, there is no evidence of FM correlations in the neutron scattering structure factor. Also, the derived witness spin interactions mediated by $Z_2$ or U(1) quantum spin liquids (see below) are purely AF, and therefore are incompatible with a FM cluster formation scenario.

6. *Spin-wave mediating the witness spin interaction*

Spin-wave-like excitations could be seen in an insulating spin glass, for example in $Eu_xSr_{1-x}S$ [75], and the method detailed in ref. 41 for spin-wave mediated nuclear spin coupling in herbertsmithite could potentially also apply to witness spin coupling. To account for this picture, we modelled spin-wave-mediated witness spin interactions using the Kondo-Yamaji Green's function decoupling of ref. 41. The interaction oscillates in sign and decays as $1/r^2$, however, it decays as $1/r$ along high-symmetry directions. Such a slow decay is inconsistent with the experimental structure factor, as shown in Extended Data Figs. 7b,c.

Furthermore, spin waves or magnons have never been identified in herbertsmithite, even by inelastic neutron scattering at 50 mK[17,27,28], thus excluding a spin-wave mediation scenario.

## B. General Impurity Spin Effects

1. Role of impurities in herbertsmithite and possible interaction with quantum spin liquids

The quasi-free "impurity" spin contribution has been known to exist in herbertsmithite since the pioneering studies[12,13,14,20,56], and there has been discussions on how it could affect the physics of herbertsmithite.

While impurities and kagome spins tended to be treated as a separate contribution in experimental analyses, the interplay between them gradually became regarded as relevant. To list a few examples, there has been discussions on a possible coupling of impurity spins to spinons to form a Kondo type ground state[23,29], a possible Jahn–Teller driven distortion leading to a displacement of the six adjacent oxygens to induce an in-plane staggered magnetic response[24], and impurity-scattering of thermal carriers[54].

Ref. 24 points out that impurity spins could offer a handle to probe the physics of the kagome layer. Our work concretely formulates and executes this concept that spin noise dynamics of "impurity" witness spins can probe the kagome layer physics, and quantitatively demonstrates that this approach is successful in deducing spinon-mediated interaction mechanisms via the kagome quantum spin liquid.

2. Spin glass transitions in spin liquid candidates

There are several materials classes that are hypothesized to be spin liquids yet exhibit a spin-glass transition at low temperatures. Here we list an example from each of Cu, Yb, Ir, Cr, Fe, and Gd-based compound families.

- $Cu_3V_2O_7(OH)_2{\cdot}2H_2O$: $Cu^{2+}$ ions with $S = 1/2$ form a distorted kagome lattice in a monoclinic structure. It exhibits a spin freezing at 1.2 K [76], or at 60 mK in another sample[57], despite Curie-Weiss temperature of $\theta_{\mathrm{CW}} = -115$ K. The transition is proposed to arise from freezing of defect-induced moments within the kagome plane[57].
- $YbMgGaO_4$ : $Yb^{3+}$ (spin-orbital $J_{\mathrm{eff}} = 1/2$) form a triangular lattice, with site-mixing disorder of non-magnetic $Mg^{2+}$ and $Ga^{3+}$ causing random local $Yb^{3+}$-environment distortion[77]. It exhibits a spin freezing at $\sim 0.1$ K ($\theta_{\mathrm{CW}}$ of $-10$ K order), suggested to be driven by disorder and frustration[78].
- $Na_4Ir_3O_8$: $Ir^{4+}$ (spin-orbital $J_{\mathrm{eff}} = 1/2$ ) form a hyperkagome lattice, exhibiting a disordered magnetic freezing of all $Ir^{4+}$ sites at $\sim 7$ K [79] ( $\theta_{\mathrm{CW}} \sim -650$ K [80]). The inhomogeneous magnetic ground state is proposed to be driven either by disorder inherent to the creation of the hyperkagome lattice or via quantum fluctuations[81].
- $SrCr_{9p}Ga_{12-9p}O_{19}$: $Cr^{3+}$ ($S = 3/2$) form a kagome-triangular-kagome trilayer lattice[82]. It shows a spin-glass-like transition at $\sim$3 K ( $\theta_{\mathrm{CW}} \sim -500$ K) in a wide range of stoichiometry $p > 0.6$, which occurs in bulk but not due to isolated impurities[83]. The transition has been given different interpretations, for example, freezing of magnetic defects localized around spin vacancies[83], or a spin jam state caused by quantum fluctuations in a disorder-free lattice[84].

- $(H_3O)Fe_3(SO_4)_2(OH)_6$: $Fe^{3+}$ ($S = 5/2$) form a kagome lattice, showing a spin glass transition at 17 K ($\theta_{CW} \sim -1200$ K)[85]. The transition is proposed to be due to a coherent anisotropic distortion of oxygen octahedra that generates in-plane anisotropy, not from random disorder[86].
- $Gd_3Ga_5O_{12}$: $Gd^{3+}$ ($S = 7/2$) form a hyperkagome lattice, showing a spin glass transition at 0.14 K ($\theta_{CW} \sim -2$ K) [64]. The transition is proposed to be due to highly frustrated geometry of the magnetic lattice[64], or a mixture of a spin-liquid state with a set of rigid magnetic pieces nucleated around impurity centers[87].

Importantly, for all these QSL candidates, the microscopic spin physics is quite different from that of herbertsmithite. The majority of these compounds have spin $S \geq 3/2$ or spin-orbital $J_{\text{eff}} = 1/2$ each of which are quantum-mechanically distinct nature from pure $S = 1/2$ in herbertsmithite. Further, it is the spins on the frustrated lattice hypothesized to become a quantum spin liquid that exhibit a spin glass transition in all these materials classes, not the "impurity" or witness spins.

Thus, herbertsmithite appears quite unique in that no spin glass transition occurs for the spins of the kagome planes and, instead, kagome-mediated witness spin interactions generate a spin glass of the "impurity" spins. This is highly distinct from the phenomena reported in refs. 57,64,76-87, and provides a unique new perspective on the $T \to 0$ ground state of herbertsmithite in which the kagome spins remain a candidate quantum spin liquid.

3. Spin noise in classical spin glasses

Magnetic noise has been used previously to study conventional spin glass compounds[88] and Josephson junction arrays[89,90], each of which generates strong magnetization signals. $1/f^{\alpha}$-type power spectral density is commonly observed in different classical spin glasses[88], while the quantitative characteristics vary among compounds. For example, in $Eu_{0.4}Sr_{0.6}S$, the spin noise power continues to grow below the glass transition[91]. Another example is $CdIn_{0.3}Cr_{1.7}S_4$ where the power exponent decreases above the glass transition at 16.6 K, but then starts increasing again above 17.5 K [92,93]. These observations differ from herbertsmithite's noise, naturally because the microscopic physics in these systems is unrelated to that of the witness spin interactions in a quantum spin liquid. Moreover, the far more technically advanced systems developed recently, including achievement of sensitivity optimization reaching $fT/\sqrt{Hz}$ clean noise floor and millikelvin spin-noise measurements, have been introduced specifically to probe spin liquids[30,31,33].